\documentclass[11pt,a4paper]{article}
\usepackage{jheppub}
\usepackage{mathrsfs}
\usepackage{amsfonts}
\usepackage{setspace}
\usepackage{cellspace}
\usepackage{amsmath,amssymb,bm}
\usepackage[colorlinks=true,linkcolor=blue]{hyperref}
\usepackage{xcolor}
\usepackage{epsfig}
\usepackage{slashed}
\usepackage{caption}
\usepackage{hhline,multirow,tabularx}  
\usepackage{dcolumn}    
\usepackage{url}        
\usepackage{braket}     
\renewcommand{\bra}[1]{\left<#1\left|}
\renewcommand{\ket}[1]{\right|#1\right>}
\preprint{CNF-UMD-2020}

\title{Novel twist-three transverse-spin sum rule for the proton and related generalized parton distributions}

\author[a]{Yuxun~Guo}
\author[a,b]{,Xiangdong~Ji}
\author[b]{and Kyle~Shiells}
\affiliation[a]{University of Maryland,\\ College Park, MD 20742 USA}
\affiliation[b]{Center for Nuclear Femtography,\\ 1201 New York Ave., NW, Washington DC, 20005, USA}
\emailAdd{yuxunguo@umd.edu}
\emailAdd{xji@umd.edu}
\emailAdd{kshiells@sura.org}
\abstract{
we derive a new twist-3 partonic sum rule for the transverse spin of the proton, which involves the well-know quark spin structure function $g_T(x)=g_1(x)+g_2(x)$, the less-studied but known transverse gluon polarization density $\Delta G_T(x)$, and quark and gluon canonical orbital angular momentum densities associated with transverse polarization. This is the counter part of the sum rule for the longitudinal spin of the proton derived by Jaffe and Manohar previously. We relate the partonic canonical orbital angular momentum densities to a new class of twist-3 generalized parton distribution functions which are potentially measurable in deep-virtual exclusive processes. We also discuss in detail an important technicality related to the transverse polarization in the infinite momentum frame, i.e., separation of intrinsic contributions from the extrinsic ones. We apply our finding to the transverse-space distributions of partons, angular momentum, and magnetic moment, respectively, in a transversely polarized proton.
}
\date{\today}
\begin{document}
\maketitle

\section{Introduction}

The spin structure of the proton or nucleon has been an important topic in hadronic physics for over the past 30 years~\cite{report2011,Aidala:2012mv,Deur:2018roz,Ji30y}.  This was spurred on after the so-called ``spin crisis" triggered by the European Muon Collaboration experiment \cite{Ashman:1987hv}, which showed that the quark spin contribution to the proton's spin was consistent with zero.  A first-principles explanation as to why the nucleon has spin $\frac{\hbar}{2}$ is a subtle task and has garnered much attention over the years~\cite{Leader2014}. In physics, sum rules express a total quantity as a sum of all its individual sources. The implementation of this general idea is illuminating, as it allows one to completely account for all of the contributions to a certain physical quantity without leaving out anything important. Angular momentum (AM) sum rules have been derived and studied for the proton since the 1990's. The sum rule derived for longitudinal spin of the nucleon in ref. \cite{JaffeMan90} has a simple partonic interpretation in the infinite momentum frame and light-front gauge. On the other hand, the covariant spin sum rule derived in ref. \cite{Ji1997} is valid for any reference frame, and for both longitudinal and transverse polarizations.

With regards to testing AM sum rules, progress has been made to evaluate various contributions to the nucleon's AM using quantum chromodynamics (QCD) on the lattice (see for example \cite{Deka2015,Lin2018,Alexandrou:2020sml,Yang:2016plb,Ji30y}).  Another avenue is through the measurement of generalized parton distributions~(GPDs).
Although GPDs were initially defined for mainly technical reasons~\cite{Muller1994}, eventually they were realized as a powerful way to reveal new information about the internal structure of the nucleon \cite{Ji1997,DIEHL2003,BELITSKY2005,Burkardt2003,Polyakov:2018zvc}.  They are a generalization to the simpler case of parton distribution functions (PDFs), and can be probed in deeply-virtual exclusive processes~\cite{Ji1997DVCS,Collins1997}.
AM is expressible in terms of spatial moments of the energy momentum tensor (EMT) which involves off-forward matrix elements of the nucleon, and hence GPDs naturally encode information about the nucleon's AM. 
Therefore one can relate AM contributions to PDFs and GPDs, and both are quantities which can be constrained by measurements. Consequently, spin programs in semi-inclusive deep inelastic scattering and polarized p-p collisions have been launched to measure the quark spin contributions of different flavours as well as the gluon helicity~\cite{Aidala:2012mv,Deur:2018roz}.  
Although GPDs have been studied for over 20 years, there has been a substantial revival in their interest recently as the $12~\text{GeV}$ physics program at Jefferson Lab comes to fruition and Electron-Ion Collider (EIC) is in full planning \cite{Accardi:2012qut}. A new generation of high energy experiments can potentially constrain them unlike ever before.      

The covariant sum rule that gives the leading contribution to AM in terms of light front power counting has been well studied~\cite{Ji1997}, which is shown to have a simple partonic interpretation in transversely-polarized nucleons~\cite{Ji2012,Ji20132}. However,
much controversy has been generated in the literature when transverse AM is involved ~\cite{Leader:2012md,Harindranath:2012wn,Ji:2013tva,Leader:2011cr,HATTA20121}. The issue has been recently resolved through
an explicit construction of the sum rule
using the transverse-AM expectation
value in the transversely polarized nucleon state and by the elimination of the center-of-mass contribution~\cite{Ji2020}. 
In addition to that, there is also a sub-leading
contribution in the infinite momentum frame. 
It is well-known that there is a $g_2(x)$
distribution related to quark transverse spin 
which can be measured through
deep-inelastic scattering~\cite{Jaffe:1990qh}. 
Furthermore, there is a gluon transverse polarization distribution introduced in ref.~\cite{Ji:1992eu} and studied in detail in 
ref.~\cite{Hatta2013}. Although alluded to in ref. \cite{Hatta2013}, a complete sum rule of transverse AM at the sub-leading level has not yet been derived.

In this paper, we follow the approach in 
ref. \cite{Ji2020}, deriving a new partonic  
transverse-spin sum rule for the nucleon which involves the familiar distributions $g_2(x)$ and
$\Delta G_T(x)$. Furthermore, it involves
the partonic transverse  OAM
which requires the introduction of new twist-three GPDs. The new sum rule can be viewed as the partonic counter part of the Jaffe-Monahar sum rule for the longitudinal spin.  Also crucial in this derivation is the operation of isolating intrinsic sources of AM over the sources associated with the nucleon's center-of-mass motion and we explain how this is achieved in the paper.

\begin{figure}[t]
    \centering
    \includegraphics[width=0.65\textwidth]{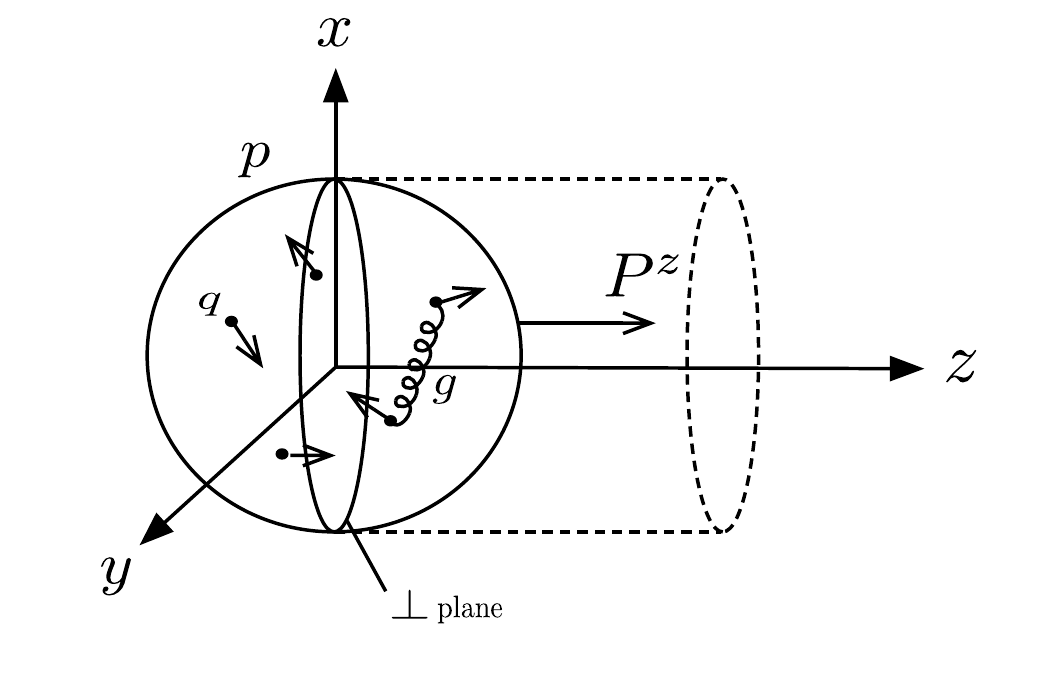}
    \caption{A proton moving along the $z$-axis
    with an intrinsic AM in both the transverse and longitudinal directions due to the internal dynamics of its quark and gluon constituents (both their motion and their spin orientation).}
    \label{fig:my_label}
\end{figure}

GPDs have been shown useful to map the transverse-plane distributions of partons~\cite{Burkardt2003}. The transverse polarization effects on these distributions already come in at the twist-2 level~\cite{Burkardt2005}. However, they often mix in the spurious contributions from the center-of-mass motions, which
has not been discussed properly in the literature before.
In this paper, we also consider how to separate out these contributions from the intrinsic ones. We show the examples
of the parton momentum, AM, and magnetic moment densities. 

The structure of the paper is as follows.  We first recommend the reader follow Appendix \ref{defam} for a refresher on how the AM is defined in quantum field theory.  Section \ref{sec2} is dedicated to defining all of the GPDs we will be considering, along with some of their properties.  Some additional details of our definition is explained in Appendix \ref{paramapdx} for the interested reader.  Section \ref{sec3} gives a detailed review of the longitudinal AM of the nucleon, which has been re-cast in terms of our defined GPDs. Section \ref{sec4} gives a detailed analysis of the sum rules for a transversely-polarized nucleon, including new results.  In section \ref{sec5} this analysis is applied to transverse-space parton distributions as well as the magnetic moment of the nucleon.  Finally we draw some conclusions and outlook in section \ref{sec6}.

\section{Twist-2 and twist-3 GPDs related to EMT}
\label{sec2}
We start in this section with a comprehensive introduction to GPDs useful for studying the matrix elements of QCD EMT and its generalization to light-front correlators. 
GPDs should be thought of as probability amplitudes for removing and adding quarks(antiquarks) or gluons from and to the nucleon in exclusive hard scattering processes. One of the simplest GPDs involving quark fields can be expressed as
\begin{equation}
\label{qqGPD}
\int_{-\infty}^{\infty}\frac{\text{d}\lambda}{2\pi}e^{i\lambda x}\bra{P',S'}\bar\psi\left(-\frac{\lambda n}{2}\right) W\left(-\frac{\lambda n}{2},\frac{\lambda n}{2}\right)\Gamma \psi\left(
\frac{\lambda n}{2}\right) \ket{P,S}\ ,
\end{equation}
where $W\left(y^\mu, x^\mu\right) \equiv P \exp \left(-i g \int_{x}^{y} d z^\mu A_\mu\left(z\right)\right)$ is the gauge link required to keep the expression gauge invariant  and $\Gamma$ is a certain Dirac matrix. The hadron states used in this paper are normalized as follows
\begin{equation}
\begin{split}
\braket{P',S'|P,S}=2E_{P}(2\pi)^3\delta^{(3)}\delta_{S,S'}(\boldsymbol P' -\boldsymbol P)\ .
 \end{split}
\end{equation}
All matrix elements will be normalized with a factor of $1/\braket{P,S|P,S}$ which will be implicit for simplicity. To aid in the analysis of expressions like eq.~(\ref{qqGPD}), the frame in which the average momentum of initial and final states, $\bar P^\mu\equiv \frac{1}{2}(P^\mu+P'^\mu)$, is collinear such that $\bar P^\perp =0$, is chosen. The average momentum can be expressed with light cone vectors as $\bar P^\mu = p^\mu +\frac{1}{2}\tilde M ^2 n^\mu$ where $p^\mu$ and $n^\nu$ are chosen in the $+$ and $-$ directions respectively defined by the light-cone coordinates $x^{\pm}=(x^0\pm x^3)/\sqrt{2}$, and satisfy $p \cdot n=1$, while $\tilde {M}^2\equiv M^2-\Delta^2/4$ and $\Delta^\mu\equiv P'^\mu-P^\mu$. In terms of 4-vectors, the two light cone vectors are $p^\mu=\frac{1}{\sqrt{2}}(\mathcal P,0,0,\mathcal P)$ and $
n^\mu=\frac{1}{\sqrt{2}}(1/\mathcal P,0,0,-1/\mathcal P)$ with $\mathcal P$ specifying the coordinate system, usually chosen to be $\bar P^+$. The momentum transfer squared is defined as $t\equiv \Delta^2$ and $S^\mu$ is the polarization vector of the particle satisfying $S\cdot P =0$ and normalized according to $S^2=-1$.
\subsection{Quark GPDs with bilinear fields}
In order to study the properties of GPDs in more details, it will be helpful to seek a set of independent functions to parameterize the matrix element.  Consider the matrix element of bi-local quark operators defined as the following
\begin{equation}\label{vectorGPD}
\begin{split}
\int_{-\infty}^{\infty}\frac{\text{d}\lambda}{2\pi}e^{i\lambda x}\bra{P',S'}\bar\psi\left(-\frac{\lambda n}{2}\right)W_{-\frac{\lambda}{2},\frac{\lambda}{2}}\Gamma\psi\left(
\frac{\lambda n}{2}\right)  \ket{P,S}&\\
=\bar u(P',S') &\mathcal F_{q,\Gamma}(x,\bar P,\Delta,n) u(P,S)\ ,
\end{split}
\end{equation}
where $\mathcal F_{q,\Gamma}(x,\bar P,\Delta,n)$ indicates a set of independent and complete functions of different Dirac structures preserving all the symmetries such as the Lorentz and parity symmetry and $q$ represents the quark flavor. One of the most well-known parameterizations for (\ref{vectorGPD}) is that of $\Gamma=\gamma^+$, in which one has~\cite{Ji1997}
\begin{equation}
\label{gamma+exp}
\mathcal F_{q,\gamma^+}=H_q(x,t,\xi) \gamma^+ + E_q(x,t,\xi) \frac{i\sigma^{+\alpha} \Delta_\alpha}{2M}\ ,
\end{equation}
where $\xi \equiv -\Delta \cdot n/2$ is the only inner product independent of the other scalars.  GPDs defined with $\Gamma=\gamma^+$ are an example of twist-2 GPDs, which dominate in most high energy scattering observables. Dynamically, operators of twist-$t$ grow like $Q^{2-t}$ as the momentum transfer $Q$ increases, so the higher order terms will be twist-3, twist-4, etc. Another definition of twist is from the operators, where
\begin{equation}
\text{twist}(t)=\text{Dimension of the operator}(d)-\text{Lorentz spin}(s)\ .
\end{equation}
This is consistent with the definition based on power counting. For instance for $\Gamma=\gamma^+$, we have $t=3-1=2$ for a dimension-3 and spin-1 operator, which is the same as the twist from powering counting. 

One important application of GPDs is that one can constrain the gravitational form factors from GPDs defined in this way, and the parameterization above leads to the twist-2 sum rule of quark AM as in ref. \cite{Ji1997}
\begin{equation}
J_{q}=\int_{-1}^{1} \frac{x}{2}\left(H_q(x,0,0)+E_q(x,0,0)\right)\text{d} x\ .
\end{equation}
The light front momentum fraction $x$ should always take the value from $-1$ to $1$ by definition, which applies to the integral of $x$ as well. Thus it will be suppressed in future equations. In order to study the higher twist contributions of GPDs to the AM, a more general parameterization with $\Gamma=\gamma^\mu$ will be needed as will be discussed later. Using the techniques highlighted in Appendix \ref{paramapdx}, one can write the following parameterization for $\mathcal F_{\gamma^\mu}$:
\begin{equation}\label{gam}
\begin{split}
\mathcal F_{q,\gamma^\mu}=&\gamma^\mu H'_q(x,t,\xi)  + \frac{i\sigma^{\mu\alpha} \Delta_\alpha}{2M} E'_q(x,t,\xi) \\
&+\frac{\Delta^\mu}{M}G'_{q,1}(x,t,\xi)+\Delta^\mu\slashed n G'_{q,2}(x,t,\xi)+ \frac{i\sigma_{\perp}^{\mu\alpha}\Delta_\alpha}{2M}G'_{q,3}(x,t,\xi)\\
&+ Mi\sigma^{\mu\nu}n_\nu G'_{q,4}(x,t,\xi)\\
&+ M n^\mu F'_{q,1}(x,t,\xi) + M^2 n^\mu \slashed n F'_{q,2}(x,t,\xi)\ ,
\end{split}
\end{equation}
where $\slashed n=\gamma^\mu n_\mu$ and $\sigma_{\perp}^{\mu\alpha}$ are non-zero for $\mu$ in the transverse direction only. Although this expression is manifestly Lorentz invariant, the fact that each component has a covariant $\mu$ index makes it difficult to separate contributions of different twist. To make the twist counting more apparent, one could reparameterize eq.~(\ref{gam}) into a twist-separated form as 
\begin{equation}\label{gamproj}
\begin{split}
\mathcal F_{q,\gamma^\mu}=&\gamma_{+}^\mu H_q(x,t,\xi)  + \frac{i\sigma_{+}^{\mu\alpha} \Delta_\alpha}{2M} E_q(x,t,\xi) \\
&+\frac{\Delta^\mu_{\perp}}{M}G_{q,1}(x,t,\xi)+\Delta^\mu_\perp\slashed n G_{q,2}(x,t,\xi)+ \frac{i\sigma_{\perp}^{\mu\alpha}\Delta_\alpha}{2M}G_{q,3}(x,t,\xi)\\
&+ M i\sigma_\perp^{\mu\nu}n_\nu G_{q,4}(x,t,\xi)\\
&+ M n_-^\mu F_{q,1}(x,t,\xi) + M^2 n_-^\mu \slashed nF_{q,2}(x,t,\xi)\ ,
\end{split}
\end{equation}
where each $\mu$ comes with a subscript indicating that it is non-zero when $\mu$ takes the value of the subscripts only.  This form can be derived by simply projecting $\mathcal F_{q,\gamma^\mu}$ to the different directions $+,\perp,-$, and the relation between the primed and unprimed GPDs is not particularly useful as only eq.~(\ref{gamproj}) will be used from this point on. The result shown in this way can be easily proved equivalent to the GPD parameterization in ref. \cite{Meisner2009} as will be shown later.

Both this parameterization and the one in ref. \cite{Meisner2009} indicates that there are two twist-2 GDPs for $\Gamma=\gamma^+$, four twist-3 GPDs for $\Gamma=\gamma^\perp$ and two twist-4 GPDs for $\Gamma=\gamma^-$. The counting of the total number of independent terms in these expansions can be achieved by counting the physical degrees of freedom as stated in ref. \cite{Ji1997DVCS,Diehl2001,Ji2001counting,HAGLER2004164,zhang2005}. For each operator $\hat O$ the matrix element $\bra{P',S'}\hat O\ket{P,S}$
has in general 4 matrix elements, each with the initial and final proton states taking different helicities. However the parity symmetry relates half of them to the others, so for each independent operator, or in this case for each free index in the operator, 2 independent terms are needed to cover all possible matrix elements.
For instance, in the twist-3 case, one finds 
\footnote{In ref. \cite{Kiptily2002} and ref. \cite{PENTTINEN2000}, equivalent twist-3 GPDs are defined but with $G_2$ and $G_3$ switched. Here we take the notation consistent with ref. \cite{PENTTINEN2000}.  }
\begin{equation}
\label{vectorperpexp}
\begin{split}
\mathcal F_{q,\gamma^\perp}= 
&\frac{\Delta^{\perp}}{M}G_{q,1}(x,t,\xi)+\Delta^\perp\slashed nG_{q,2}(x,t,\xi)+ \frac{i\sigma^{\perp\rho}\Delta_\rho}{2M}G_{q,3}(x,t,\xi)\\
&+ M i\sigma^{\perp\rho}n_\rho G_{q,4}(x,t,\xi)\ ,
\end{split}
\end{equation}
that has 4 independent parameters which agrees with the counting rule above (2 for each transverse direction). Another choice of parameterization in the literature is using the structure in the form of $\mathcal F_{q,\gamma^\perp}\sim \gamma^{\perp} \gamma_5 G(x,t,\xi)$ \cite{Kiptily2002,PENTTINEN2000,Ji2012,Aslan:2018zzk}, and their comparison is discussed in Appendix \ref{otherparam}. This analysis of counting independent matrix elements will be very useful in examining some of the more complicated parameterizations in the following sections.
\subsection{Quark GPDs with three fields}
\label{3fieldparam}
More complicated GPD parameterizations for $\Gamma$ assuming some general Dirac matrix can be done similarly. For instance in ref. \cite{Meisner2009,Diehl2001}, GPDs with $\Gamma=\sigma^{\mu\nu}$ have been parameterized. Another structure defined as
\begin{equation}
\label{qqDgpd}
\int \frac{\text{d}\lambda \text{d}\mu}{(2\pi)^2}e^{i\frac {\lambda} {2} (x+y)+i\mu(y-x)}\bra{P',S'}\bar\psi\left(-\frac{\lambda n}{2}\right) W_{-\lambda/2,\mu}\gamma^\alpha i\overleftrightarrow{D}^\beta(\mu n) W_{\mu ,\lambda/2 }\psi\left(
\frac{\lambda n}{2}\right) \ket{P,S}\ ,
\end{equation}
where the double arrow covariant derivative is defined as $\overleftrightarrow{D}=\frac{1}{2}\left(\overrightarrow{D}-\overleftarrow{D}\right)$ with $\overrightarrow{D}=\overrightarrow{\partial}+igA$ and $\overleftarrow{D}=\overleftarrow{\partial}-igA$, are called D-type GPDs \cite{Ji20132} which are related to the quark EMT as will be shown. Another form,
\begin{equation}
\label{Ftypedef}
\int \frac{\text{d}\lambda \text{d}\mu}{(2\pi)^2}e^{i\frac {\lambda} {2} (x+y)+i\mu(y-x)}\bra{P',S'}\bar\psi\left(-\frac{\lambda n}{2}\right) W_{-\lambda/2,\mu}\gamma^\alpha  ig  G^{+\beta}(\mu n) W_{\mu ,\lambda/2 }\psi\left(
\frac{\lambda n}{2}\right) \ket{P,S}\ ,
\end{equation}
called the F-type GPDs which can be related to the transverse color Lorentz force on quarks\cite{Aslan:2019jis} are of interest when AM is concerned. The D- and F-type GPDs are accompanied with two free indices $\alpha,\beta$ similar to the $\Gamma=\sigma^{\mu\nu}$ case, but no specific symmetry with exchanging the indices is required, unlike $\sigma^{\mu\nu}$ which is anti-symmetric. Consequently, a more general parameterization is needed. Also, the D- and F-type GPDs are defined with a correlation of 3 fields separated on the light front, which makes them even more complicated. As will be shown, these GPDs will suffice for studying the AM.

Attempting to parameterize those GPDs with a set of complete and independent functions, eq.~(\ref{qqDgpd}) and eq.~(\ref{Ftypedef}) can be expressed as
\begin{equation}
\text{D/F-type quark GPD} =\bar u(P',S') \mathcal F^{\alpha\beta}_{q,D/F}(x,y,\bar P,\Delta,n) u(P,S)\ .
\end{equation}
Then the following expression can provide a parameterization (See Appendix \ref{paramapdx})
\begin{equation}
\begin{split}
\label{tensorGPDexp}
\mathcal F_{q,D}^{\alpha\beta}=&M g_\perp^{\alpha \beta}A_1(x,y,t,\xi) +M^2 g_\perp^{\alpha \beta} \slashed n A_2(x,y,t,\xi)+M \sigma_{\perp}^{\alpha \beta} A_3(x,y,t,\xi)\\
&+M V_S^\alpha \otimes V_S^\beta + M^2 \slashed n V_S^\alpha \otimes V_S^\beta+M V_S^\alpha \otimes V_T^\beta +M V_T^\alpha \otimes V_S^\beta\ ,
\end{split}
\end{equation}
where $V_S^\alpha$ is the linear space expanded by vectors with Dirac scalar
\begin{equation}
V_S^\alpha=\left\{\frac{\bar P^\alpha_+}{M},\frac{ \Delta_\perp^\alpha}{M},Mn_-^\alpha\right\}\ ,
\end{equation}
$V_T^\alpha$ is the linear space expanded by vectors with a Dirac tensor
\begin{equation}
V_T^\alpha=\left\{Mi \sigma_\perp^{\alpha\beta}n_\beta,\frac{ i \sigma_\perp^{\alpha\beta}\Delta_\beta}{2M} \right\}\ ,
\end{equation}
and $\otimes$ in eq.~(\ref{tensorGPDexp}) represents the direct product of the two linear spaces. Each base vector will be associated with an independent function $A_n(x,y,t,\xi)$. It can be shown that there are 32 independent structures and as a result 32 independent functions $A_n(x,y,t,\xi)$. One can check its consistency with other parameterization by taking the anti-symmetric part of $F^{\alpha\beta}_{q,D/F}(x,y,\bar P,\Delta,n)$. It turns out there are 12 independent functions for anti-symmetric $F^{\alpha\beta}_{q,D/F}(x,y,\bar P,\Delta,n)$ which are equivalent to the results in ref. \cite{Meisner2009}.
Though the parameterization seems to be complicated, the number of independent coefficients is consistent with the parameter counting method. Since there are $4\times 4=16$ free indices for a rank-2 tensor and $(4\times 3)/2=6$ free indices for an anti-symmetric rank-2 tensor, there should be at most 32 and 12 independent parameters for each of them respectively.
The twist-3 part of the tri-local GPDs is essential when the AM is of interest after the twist-2 terms, so consider eq.~(\ref{tensorGPDexp}) with $\alpha=+,\beta=\perp$ which gives
\begin{equation}
\label{tensor+perpexp}
    \begin{split}
       \frac{1}{\bar P^+} \mathcal F_{q,D}^{+\perp}=&\frac{\Delta^{\perp}}{M} G_{q,D,1}(x,y,\xi,t) + \Delta^{\perp}\slashed n G_{q,D,2}(x,y,\xi,t)+ \frac{i\sigma^{\perp\rho}\Delta_{\rho}}{2M} G_{q,D,3}(x,y,\xi,t)\\
        &+M i \sigma^{\perp\rho}n_{\rho} G_{q,D,4}(x,y,\xi,t)\ .
    \end{split}
\end{equation}
The parameterization of the F-type GPDs will also be defined as
\begin{equation}
\label{quarkFexp}
    \begin{split}
       \frac{1}{\left(\bar P^+\right)^2} \mathcal F_{q,F}^{+\perp}=&\frac{\Delta^{\perp}}{M} G_{q,F,1}(x,y,\xi,t) + \Delta^{\perp}\slashed n G_{q,F,2}(x,y,\xi,t)\\
        &+ \frac{i\sigma^{\perp\rho}\Delta_{\rho}}{2M} G_{q,F,3}(x,y,\xi,t)+M i\sigma^{\perp\rho}n_{\rho} G_{q,F,4}(x,y,\xi,t)\ .
    \end{split}
\end{equation}
If one instead takes $\beta=+$ inside $\mathcal{F}^{\alpha\beta}_{q,D/F}$, the GPDs will be much simpler in the light front gauge. As with $A^+=0$, $\overleftrightarrow{D}^+(\mu n)$ reduces to $\overleftrightarrow{\partial}^+$ where the $\mu$ dependence is removed and it becomes $xP^+$ up to an integration by parts. As a result, one has the following general substitution for $\beta=+$:
\begin{equation}
\begin{split}
\label{partial+subs}
\int \frac{\text{d}\lambda \text{d}\mu}{(2\pi)^2}e^{i\frac {\lambda} {2} (x+y)+i\mu(y-x)}\bar\psi&\left(-\frac{\lambda n}{2}\right) \gamma^{\alpha} i \overleftrightarrow{D}(\mu n)^{+} \psi\left(
\frac{\lambda n}{2}\right).\\
&\to x \mathcal P\delta(y-x)\int_{-\infty}^{\infty}\frac{\text{d}\lambda}{2\pi}e^{i\lambda x}\bar\psi\left(-\frac{\lambda n}{2}\right)\gamma^\alpha \psi\left(
\frac{\lambda n}{2}\right)\ ,
\end{split}
\end{equation}
where the gauge links reduce to unity in the light front gauge, while the F-type GPDs simply vanish since $G^{++}=0$. Then the off-forward matrix elements of the relations above would imply
\begin{equation}
\label{partial+subsGPD}
\mathcal F_{q,D}^{\mu +}(x,y,t,\xi) \to x\bar P^+ \delta(y-x) \mathcal F_{q,\gamma^\mu}(x,t,\xi)\ .
\end{equation}
Thus all D-type GPDs with $\beta=+$ can be expressed by the 2-field GPDs in the light front gauge, while the F-type GPDs simply vanish
\begin{equation}
\mathcal F_{q,F}^{\mu +}(x,y,t,\xi) =0\ .
\end{equation}
\subsection{Gluon GPDs with bilinear and three fields}
The gluon GPDs can be defined very similarly as the quark ones. For instance, GPDs defined with a gluon-gluon correlation can be expressed as
\begin{equation}
\label{gluonGPDdef}
\int_{-\infty}^{\infty}\frac{\text{d}\lambda}{2\pi}e^{i\lambda x}\left<P',S'\left|2\text{Tr}\left\{ G^{(\alpha i}\left(-\frac{\lambda n}{2}\right) W_{-\frac{\lambda}{2},\frac{\lambda}{2}} G_{\;\;\;\;i}^{\beta)}\left(
\frac{\lambda n}{2}\right)\right\} \right|P,S\right>\ ,
\end{equation}
where $()$ in the indices indicate a symmetrization. This expression can be parameterized in much the same way as in eq.~(\ref{tensorGPDexp}), with the only new constraint being the need to symmetrize the $\alpha,\beta$ indices. With the expansion of eq.~(\ref{gluonGPDdef}) as 
\begin{equation}
\text{Gluon GPD} =\bar u(P',S') \mathcal F^{\alpha\beta}_{g}(x,\bar P,\Delta,n) u(P,S)\ ,
\end{equation}
the following parameterization is chosen
\begin{equation}
\begin{split}
\mathcal F_{g}^{\alpha\beta}=&M g_\perp^{\alpha \beta}A_1(x,t,\xi) +M^2 g_\perp^{\alpha \beta} \slashed n A_2(x,t,\xi)\\
&+M V_S^{(\alpha} \otimes V_S^{\beta)} + M^2 \slashed n V_S^{(\alpha} \otimes V_S^{\beta)}+M V_S^{(\alpha} \otimes V_T^{\beta)}\ .
\end{split}
\end{equation}
There are 20 independent parameters which is again consistent with the counting method of independent matrix elements. The twist-2 part of the above parameterization with $\alpha=\beta=+$ can be written as
\begin{equation}
\begin{split}
\label{gluonparameter1}
\frac{1}{\bar P^+}\mathcal F_{g}^{++}=\gamma^+  xH_g(x,t,\xi)+\frac{i\sigma^{+\alpha}\Delta_\alpha}{2M} x E_g(x,t,\xi)\ ,
\end{split}
\end{equation}
in which a Gordon identity was used and a factor of $x$ included so that the parameterization matches with the common result. For the twist-3 GPDs with $\alpha=+, \beta=\perp$, one has
\begin{equation}
\begin{split}
\label{gluonparameter2}
\mathcal F_{g}^{+\perp}=\bar P^{(+} &\bigg{[}\frac{\Delta^{\perp)}}{M} x G_{g,1}(x,\xi,t) + \Delta^{\perp)}\slashed n x G_{g,2}(x,\xi,t)+ \frac{i\sigma^{\perp)\rho}\Delta_{\rho}}{2M} x G_{g,3}(x,\xi,t)\\
        &\;\;\;+M  i\sigma^{\perp)\rho}n_{\rho} x G_{g,4}(x,\xi,t)\bigg{]}\ ,
\end{split}
\end{equation}
where a factor of $x$ is multiplied again for consistency. The GPDs defined here are symmetric under exchange of $\alpha $ and $\beta$, so the parameterization of $\alpha=\perp, \beta=+$ will be the same.
Besides the GPDs with a gluon-gluon correlation above, three-gluon correlation functions will be needed for reasons explained later. Consider the D-type gluon GPDs defined from
\begin{equation}
\label{gluonDtypedef}
\int \frac{\text{d}\lambda \text{d}\mu}{(2\pi)^2}e^{i\frac {\lambda} {2} (x+y)+i\mu(y-x)}\left<P',S'\left|2\text{Tr}\left\{ G^{\alpha i}\left(-\frac{\lambda n}{2}\right) i D^{\beta}(\mu n) G_{\;\;i}^{+}\left(
\frac{\lambda n}{2}\right)  \right\}\right|P,S\right>\ ,
\end{equation}
and the F-type GPDs from 
\begin{equation}
\label{gluonFtypedef}
\int \frac{\text{d}\lambda \text{d}\mu}{(2\pi)^2}e^{i\frac {\lambda} {2} (x+y)+i\mu(y-x)}\bra{P',S'}\sum_{i}2\text{Tr}\left\{ G^{\alpha i}\left(-\frac{\lambda n}{2}\right) ig G^{+\beta}(\mu n) G_{\;\;i}^{+}\left(
\frac{\lambda n}{2}\right)  \right\}\ket{P,S}\ .
\end{equation}
All the gauge links are removed since these definitions are related to observables only in the light cone gauge.  Then with their parameterization written as
\begin{equation}
\label{gluondfparam}
\text{D/F-type gluon GPD} =\bar u(P',S') \mathcal F^{\alpha\beta}_{g,D/F}(x,y,\bar P,\Delta,n) u(P,S)\ ,
\end{equation}
they will follow analogously to the quark D/F-type GPDs. For instance, the twist-2 case of $\alpha=+,\beta=+$ gives us
\begin{eqnarray}
\label{gluonsub1}
\mathcal F_{g,D}^{\alpha+}(x,y,\bar P,\Delta,n)&=&x \bar P^+ \delta(y-x)\mathcal F_{g}^{\alpha+}(x,\bar P,\Delta,n)\ ,\\
\label{gluonsub2}
\mathcal F_{g,F}^{\alpha+}(x,y,\bar P,\Delta,n)&=&0\ ,
\end{eqnarray}
where these relations hold for any $\alpha$, so one has the same relation for the twist-3 GPDs with $\alpha=\perp, \beta=+$. As for the twist-3 case with $\alpha=+,\beta=\perp$, the D-type GPDs are
\begin{equation}
\label{gluon+perpDexp}
\begin{split}
       \frac{1}{\left(\bar P^+\right)^2} \mathcal F_{g,D}^{+\perp}=&\frac{\Delta^{\perp}}{M} G_{g,D,1}(x,y,\xi,t) + \Delta^{\perp}\slashed n G_{g,D,2}(x,y,\xi,t)\\
        &+ \frac{i\sigma^{\perp\rho}\Delta_{\rho}}{2M} G_{g,D,3}(x,y,\xi,t)+M i \sigma^{\perp\rho}n_{\rho} G_{g,D,4}(x,y,\xi,t)\ ,
\end{split}
\end{equation}
and the F-type GPDs are
\begin{equation}
\label{gluon+perpFexp}
\begin{split}
       \frac{1}{\left(\bar P^+\right)^3} \mathcal F_{g,F}^{+\perp}=&\frac{\Delta^{\perp}}{M} G_{g,F,1}(x,y,\xi,t) + \Delta^{\perp}\slashed n G_{g,F,2}(x,y,\xi,t)\\
        &+ \frac{i\sigma^{\perp\rho}\Delta_{\rho}}{2M} G_{g,F,3}(x,y,\xi,t)+M i \sigma^{\perp\rho}n_{\rho} G_{g,F,4}(x,y,\xi,t)\ .
    \end{split}
\end{equation}

Among our conclusions is a catalogue of twist-2 and twist-3 GPDs spread across 6 distinct types of 2- and 3-field light cone correlations. Table \ref{GPDtable} briefly summarizes the 28 GPDs defined.

\begin{table}[t]
    \centering
    \begin{tabular}{|Sc|Sc|Sc|Sc|}
    \hline
       Correlator  & Type & twist-2 GPDs & twist-3 GPDs \\
       \hline
        $\bar{\psi}\gamma^\mu \psi$ & -- & $E_q,H_q$ & $G_{q,1},\;G_{q,2},\;G_{q,3},\;G_{q,4}$\\
        \hline
        $\bar{\psi}\gamma^\alpha i D^\beta \psi$ & D & $E_q,H_q$ & $G_{q,D,1},\;G_{q,D,2},\;G_{q,D,3},\;G_{q,D,4}$ \\
        \hline
        $\bar{\psi}\gamma^\alpha G_a^{+\beta} \psi$ & F & -- & $G_{q,F,1},\;G_{q,F,2},\;G_{q,F,3},\;G_{q,F,4}$\\
        \hline
        $G_a^{(\alpha i}G_{ai}^{\beta )}$ & -- & $E_g,H_g$ & $G_{g,1},\;G_{g,2},\;G_{g,3},\;G_{g,4}$\\
        \hline
        $G_a^{\alpha i}iD^\beta G_{ai}^{+}$ & D & $E_g,H_g$ & $G_{g,D,1},\;G_{g,D,2},\;G_{g,D,3},\;G_{g,D,4}$\\
        \hline
        $\text{Tr}\left\{G^{\alpha i}G^{+\beta} G_{\;\;i}^{+}\right\}$ & F & -- & $G_{g,F,1},\;G_{g,F,2},\;G_{g,F,3},\;G_{g,F,4}$\\
        \hline
    \end{tabular}
    \caption{A list of twist-2 and twist-3 GPDs for 6 different light cone correlators.  Each field in the 1st column is separated on the light cone, and implicitly separated by a Wilson line.  $G_{q/g,i}$ are functions of the variables $(x,\xi,t)$, while $G_{q/g,D/F,i}$ are functions of $(x,y,\xi,t)$.}
    \label{GPDtable}
\end{table}

\subsection{Constraints on GPDs from discrete symmetry}
The matrix elements defined above must also satisfy discrete symmetries such as those of parity and time reversal invariance. Also, the result should be valid under Hermitian conjugation. These conditions can relate them into constraints of the expansion coefficient functions themselves. The derivation is somewhat tedious and only the results are shown here for future convenience (see for instance \cite{BELITSKY2005,Diehl2001}). For the relevant coefficients in eq.~(\ref{vectorperpexp}) and eq.~(\ref{gluonparameter2}), the Hermiticity condition combined with time reversal symmetry require that
\begin{align}
G_{q/g,3}^*(x,t,-\xi)&=G_{q/g,3}(x,t,\xi)\ ,\\
G_{q/g,3}(x,t,-\xi)&=G_{q/g,3}(x,t,\xi)\ ,\\
G_{q/g,j}^*(x,t,-\xi)&=-G_{q/g,j}(x,t,\xi)\ ,\\
G_{q/g,j}(x,t,-\xi)&=-G_{q/g,j}(x,t,\xi)\ ,
\end{align}
with $j=\{1,2,4\}$.
Notice that under both Hermitian conjugation and time reversal, the initial and final states will be exchanged which means $\Delta \to -\Delta$ and $\xi \to -\xi$. The equations above simply imply that $G_{q/g,i}$ for $i=\{1,2,3,4\}$ are all real. Meanwhile, $G_{q/g,3}$ is an even function of $\xi$, and $G_{q/g,j}$ with $j=\{1,2,4\}$  are odd functions of $\xi$, so in the forward limit: $G_{q/g,j}(x,t=0,\xi=0)=0$.
The discrete symmetry of D-type and F-type GPDs can be derived as well and those GPDs are functions of both $x$ and $y$ which makes it more complicated. For the D-type 3-field GPDs defined in eq.~(\ref{tensor+perpexp}) and eq.~(\ref{gluon+perpDexp}), the Hermiticity condition combined with the time reversal symmetry gives
\begin{align}
G_{q/g,D,3}^*(x,y,t,-\xi)&=G_{q/g,D,3}(y,x,t,\xi)\ ,\\
G_{q/g,D,3}(x,y,t,-\xi)&=G_{q/g,D,3}(y,x,t,\xi)\ ,\\
G_{q/g,D,j}^*(x,y,t,-\xi)&=-G_{q/g,D,j}(y,x,t,\xi)\ ,\\
G_{q/g,D,j}(x,y,t,-\xi)&=-G_{q/g,D,j}(y,x,t,\xi)\ ,
\end{align}
with $j=\{1,2,4\}$. Again all the GPDs defined here are real. Also, if one expands those GPDs into powers of $\xi$ via 
\begin{align}
f(x,y,t,\xi)=\sum_{n=0}^{\infty}f^{(n)}(x,y,t)\frac{\xi^n}{n!}\ ,
\end{align}
then it follows that all the $G_{q/g,D,3}^{(n)}(x,y,t,-\xi)$ are symmetric when $n$ is even and anti-symmetric when $n$ is odd which is the opposite for $G_{q/g,D,j}^{(n)}(x,y,t,-\xi)$. 
There remains the F-type GPDs for which the constraints can be derived in the same way. Combining the Hermiticity condition and the time reversal symmetry one has
\begin{align}
G_{q/g,F,3}^*(x,y,t,-\xi)&=-G_{q/g,F,3}(y,x,t,\xi)\ ,\\
G_{q/g,F,3}(x,y,t,-\xi)&=-G_{q/g,F,3}(y,x,t,\xi)\ ,\\
G_{q/g,F,j}^*(x,y,t,-\xi)&=G_{q/g,F,j}(y,x,t,\xi)\ ,\\
G_{q/g,F,j}(x,y,t,-\xi)&=G_{q/g,F,j}(y,x,t,\xi)\ ,
\end{align}
with $j=\{1,2,4\}$. These relations indicate that $ G_{q/g,F,3}^{(n)}(x,y,t,-\xi)$  are anti-symmetric when $n$ is even and symmetric when $n$ is odd and the opposite applies to $ G_{q/g,F,j}^{(n)}(x,y,t,-\xi)$ with $j=\{1,2,4\}$. 

\subsection{Connection to the previous definitions}

The GPDs defined in the sections above can be related to those of Mei\ss ner \textit{et al.} \cite{Meisner2009}, which are also used extensively in a more recent paper \cite{Kriesten:2019jep} on deeply virtual compton scattering (DVCS) cross sections. To start the comparison, matrix element in ref. \cite{Meisner2009} that gives the GPDs can be written as
\begin{equation}
    F_{\lambda\lambda'}^{[\Gamma]}(\bar{P},x,\Delta,N)=\frac{1}{2}\int \frac{dz}{2\pi} e^{ik\cdot z} \langle p',\lambda'| \bar{\psi}(-z/2)\Gamma W\left(-\frac{z}{2},\frac{z}{2}\right)\psi(z/2)|p,\lambda\rangle_{z^+=\vec{z}_\perp=0}\ ,
\end{equation}
where the $N$ plays the same role as the negative light cone vector $n$.  Performing a change of variables $\tilde{\lambda}=z^-\bar{P}^+$, this is equivalent to the expression
\begin{equation}
 F_{\lambda\lambda'}^{[\Gamma]}=\frac{1}{2\bar{P}^+}\int \frac{d\tilde{\lambda}}{2\pi}e^{i\tilde{\lambda}x} \langle p',\lambda'\bigg{|} \bar{\psi}\bigg{(}-\frac{\tilde{\lambda}n}{2}\bigg{)}\Gamma W\left(-\frac{\tilde{\lambda}n}{2},\frac{\tilde{\lambda}n}{2}\right)\psi\bigg{(}\frac{\tilde{\lambda}n}{2}\bigg{)}\bigg{|}p,\lambda\rangle \  ,
\end{equation}
which is the left-hand side (LHS) of eq.~(\ref{vectorGPD}) with an extra factor of $1/(2\bar{P}^+)$.  This is made all the more obvious when comparing the twist-2 amplitude in eq.~(\ref{gamma+exp}) to Mei$\beta$ner \textit{et al.}'s analogous expression
\begin{equation}
    F_{\lambda\lambda'}^{[\gamma^+]} = \frac{1}{2P^+}\bigg{[} \gamma^+ H(x,\xi,t)+\frac{i\sigma^{+\Delta}}{2M}E(x,\xi,t) \bigg{]},
\end{equation}
which has the exact same decomposition. Therefore, the following general relationship can be established
\begin{equation}\label{GuoMeiz}
    F_{\lambda,\lambda'}^{[\gamma^j]}=\frac{1}{2\bar{P}^+}\mathcal{F}_{q,\gamma^j}\ ,
\end{equation}
where the momentum space spinors are suppressed and implicitly present on the right-hand side (RHS), and this in fact readily relates our twist-2 GPDs to those in ref. \cite{Meisner2009}. In particular
\begin{eqnarray}
\begin{pmatrix} H_q(x,\xi,t) \\ E_q(x,\xi,t)\end{pmatrix}=\frac{1}{2\bar{P}^+}\begin{pmatrix} H(x,\xi,t) \\ E(x,\xi,t)\end{pmatrix}\ ,
\end{eqnarray}
where our GPDs are on the LHS and Mei\ss ner \textit{et al.}'s are on the RHS. As for the twist-3 GPDs, ref. \cite{Meisner2009} gives the following amplitude
\begin{equation}
\label{Meiztw3}
\begin{split}
    F_{\lambda\lambda'}^{[\gamma^j]} &= \frac{M}{2(\bar{P}^+)^2} \Bigg{[} i\sigma^{+j}H_{2T}(x,\xi,t) + \frac{\gamma^+\Delta^j_T-\Delta^+\gamma^j}{2M}E_{2T}(x,\xi,t) \\
   & \qquad\qquad\quad+\frac{\bar{P}^+\Delta_T^j-\Delta^+\bar{P}_T^j}{M^2}\tilde{H}_{2T}(x,\xi,t) + \frac{\gamma^+\bar{P}_T^j-\bar{P}^+\gamma^j}{M}\tilde{E}_{2T}(x,\xi,t) \Bigg{]}\ .
\end{split}
\end{equation}
Recognizing the following relations which are a consequence of both simple Dirac algebra and the light cone coordinates
 \begin{eqnarray}
 \slashed{n}&=&\frac{\gamma^+}{\bar{P}^+}\ ,\\
 \frac{\Delta^+}{\bar{P}^+}&=&-2\xi\ , \\
 \frac{\sigma_\perp^{+\mu}}{\bar{P}^+}&=&\sigma_\perp^{\mu\nu}n_\nu\ ,
 \end{eqnarray}
and substituting these three relations into eq.~(\ref{Meiztw3}) gives the following expression
\begin{equation}
\begin{split}
(2\bar{P}^+)F_{\lambda\lambda'}^{[\gamma^j]} =& iM\sigma_\perp^{j\nu} n_\nu H_{2T} + \Delta_\perp^j \bigg{(} \frac{\slashed{n}}{2}E_{2T} +\frac{1}{M}\tilde{H}_{2T} \bigg{)} \\
& + \xi\gamma_\perp^j E_{2T} + \frac{2\xi}{M}\bar{P}_\perp^j\tilde{H}_{2T} +\bigg{(} \bar{P}_\perp^j\slashed{n}-\gamma_\perp^j \bigg{)}\tilde{E}_{2T}\ ,
\end{split}
\end{equation}
where all the arguments will be suppressed. In the frame where $\bar{P}_\perp=0$, it leaves one with
\begin{eqnarray}\label{Meizfinal}
(2\bar{P}^+)F_{\lambda\lambda'}^{[\gamma^j]} &=& iM\sigma_\perp^{j\nu} n_\nu H_{2T} + \Delta_\perp^j \bigg{(} \frac{\slashed{n}}{2}E_{2T} +\frac{1}{M}\tilde{H}_{2T} \bigg{)} \nonumber\\
&\;& +  \gamma_\perp^j \bigg{(} \xi E_{2T} -\tilde{E}_{2T} \bigg{)}\ .
\end{eqnarray}
The first 2 terms of eq.~(\ref{Meizfinal}) coincide with the first, second and fourth terms of eq.~(\ref{vectorperpexp}), while the third term of eq.~(\ref{vectorperpexp}) can be related to the last term in eq.~(\ref{Meizfinal}) with Gordon identity and it can be rewritten as
\begin{equation}\label{eq:147}
\bar{u}(P')\frac{i\sigma_\perp^{\mu\alpha}\Delta_\alpha}{2M}G_{q,3}(x,t,\xi)u(P) = \bar{u}(P')\gamma^{\mu}_\perp G_{q,3}(x,t,\xi)u(P)\ ,
\end{equation}
where $P^{\mu}_\perp=0$ has been used. Comparing eq.~(\ref{vectorperpexp}), eq.~(\ref{Meizfinal}) and eq.~(\ref{eq:147}), the following relations between ours and Mei\ss ner \textit{et al.}'s twist-3 GPDs can be written 
\begin{eqnarray}
\label{GPDrelation1}
E_{2T}(x,\xi,t)&=&2G_{q,2}(x,\xi,t)\ ,\\
H_{2T}(x,\xi,t)&=&G_{q,4}(x,\xi,t)\ ,\\
\tilde{E}_{2T}(x,\xi,t)&=&2\xi G_{q,2}(x,\xi,t)-G_{q,3}(x,\xi,t)\ ,\\
\label{GPDrelation4}
\tilde{H}_{2T}(x,\xi,t)&=&G_{q,1}(x,\xi,t)\ .
\end{eqnarray}
Another choice of Dirac structure has been used in the literature, see for example \cite{Ji20132,Hatta2012,Kiptily2002,Aslan:2018zzk} where the coefficient of $\bar u(P',S') i\epsilon^{\nu\rho}_{\perp}\Delta_\rho \gamma^{\perp}\gamma_5 u(P,S)$ is taken. This term is discussed in Appendix \ref{otherparam} where it is shown to be related to $\bar u(P',S') \gamma^{\perp} u(P,S)$  with eq.~(\ref{axialvectorpara}), so one has 
\begin{eqnarray}
G_{q,D,3}(x,y,t,\xi) \sim H_D^{q(3)}(x,y,t,\xi)\ ,
\end{eqnarray}
with $H_D^{q(3)}(x,y,t,\xi)$ the GPD defined in ref. \cite{Ji20132}. However, the transformation between these two different parameterizations is not as simple as the relations eq.~(\ref{GPDrelation1})-eq.~(\ref{GPDrelation4}) above due to different parameterizations being used. The forward limit of $G_{q,D,3}(x,y,t,\xi)$ also corresponds to the $\phi_D(x,y)$ defined in ref. \cite{Hatta2012} and we have
\begin{eqnarray}
G_{q,D,3}(x,y,t=0,\xi=0)=\phi_D(x,y)\ .
\end{eqnarray}
\section{Revisiting spin sum rules for longitudinal polarization}
\label{sec3}
The spin sum rules for a longitudinally polarized nucleon have been thoroughly discussed in the literature, see for instance \cite{Ji2012,Hatta2012}.  
Here these sum rules are reviewed in light of the parameterization of twist-3 GPDs in the previous section to establish a connection to the present work. In general there are 
two types of spin sum rules. One type is manifestly gauge invariant
and local~\cite{Ji1997}, and is valid in any Lorentz frame in which the helicity
is a good quantum number, 
\begin{align}
\label{jisumrule}
\frac{1}{2}\Delta q +L^z_q+J^z_g=\frac{\hbar}{2} \ ,
\end{align}
where $J^z_q = \frac{1}{2}\Delta q + L^z_q  $ is the total quark contribution in terms
of total quark helicity $\Delta q$ and quark OAM $L^z_q$, and 
$J^z_g$ is the total gluon AM. The quark contribution in AM always implies summing over all quark flavors.
The other is the Jaffe and Manohar sum rule in the infinite momentum frame and light-cone gauge~\cite{JaffeMan90},
which has simple partonic interpretations,
\begin{align}
\label{jmsumrule}
\frac{1}{2}\Delta q +\Delta G+l^z_q+l^z_g=\frac{\hbar}{2} \ , 
\end{align}
where $\Delta G$ is the total gluon helicity, and $l^z_{q,g}$ are the quark and gluon canonical AM contributions. The renormalization scale dependence is omitted for simplicity.

As discussed in Appendix \ref{defam}, the AM can be expressed in terms of the off-forward matrix elements of the EMT. For example, analogous to eq.~(\ref{Jxfinal}), $J^z_{q}$ and $J^z_{g}$ can be in general written  as
\begin{equation}
\begin{split}
\label{jznormalized}
J^z_{q,g}=\bra{P,S}J^z_{q,g}\ket{P,S}=\frac{1}{2E_P}&\left(\lim_{\Delta\to 0,S'\to S}i\frac{\partial}{\partial \Delta_x}\bra{P+\Delta,S'} T_{q,g,\text{Bel}}^{0y}(0)  \ket{P,S}\right. \\
&\left.\qquad-\lim_{\Delta\to 0,S'\to S}i\frac{\partial}{\partial \Delta_y}\bra{P+\Delta,S'} T_{q,g,\text{Bel}}^{0x}(0)  \ket{P,S}\right)\ ,
 \end{split}
\end{equation}
which allows to one to derive the matrix element of AM from the off-forward matrix elements of the EMT. 
The parameterization of the off-forward matrix element of EMT has already been done in ref. \cite{Ji1997} as 
\begin{equation}
\begin{split}
\label{Tmunuformfactors}
\bra{P'}T_{q, g,\text{Bel}}^{\mu \nu}\ket{P}=\bar{u}\left(P^{\prime}\right)&\bigg{[}A_{q, g}\left(\Delta^{2}\right) \gamma^{(\mu} \bar{P}^{\nu)}+B_{q, g}\left(\Delta^{2}\right) \bar{P}^{(\mu} i \sigma^{\nu) \alpha} \Delta_{\alpha} / 2 M\\
&\;\;+C_{q, g}\left(\Delta^{2}\right)\left(\Delta^{\mu} \Delta^{\nu}-g^{\mu \nu} \Delta^{2}\right) / M
+\bar{C}_{q, g}\left(\Delta^{2}\right) g^{\mu \nu} M\bigg{]} u(P)\ .
\end{split}
\end{equation}
In order to get the AM with eq.~(\ref{jznormalized}), expansion of above parameterization to the leading order of $\Delta$ will be needed. Taking the gamma matrices in the Dirac representation, the spinor can be written as 
\begin{equation}
u(P,\boldsymbol s)=\sqrt{P^0+M}\begin{pmatrix} \chi^{(\boldsymbol s)} \\ \frac{\boldsymbol \sigma\cdot \boldsymbol P}{P^0+M}\chi^{(\boldsymbol s)}\end{pmatrix}\ ,
\end{equation}
where $\boldsymbol \sigma$ represents the Pauli matrices, $\boldsymbol s$ is the spin vector in the rest frame and $\chi^{(\boldsymbol s)}$ satisfies
\begin{equation}
\chi^{(\boldsymbol s)\dagger }\sigma^i\chi^{(\boldsymbol s)}= \boldsymbol s^i\ .
\end{equation}
Two expansions of Dirac bilinears will be used, which are
\begin{eqnarray}
\label{spinorscalar}
\bar u(\bar P+\Delta/2,S) u(\bar P-\Delta/2,S) &=& 2M+\epsilon^{0\rho\alpha\beta} \frac{i\Delta_\rho}{\bar P^0+M}S_\alpha \bar P_\beta +O(\Delta^2)\ ,\\
\bar u(\bar P+\Delta/2,S) \sigma^{\mu\nu} u(\bar P-\Delta/2,S) &=&- 2\epsilon^{\mu\nu\alpha\beta}S_\alpha \bar P_{\beta}+O(\Delta)\ ,\label{sigmaspinor}
\end{eqnarray}
where $\epsilon^{0123}=1$ and the zeroth order expansion of the second line is enough. The expansion of eq.~(\ref{Tmunuformfactors}) can then be written as~\cite{JaffeMan90,Bakker2004}
\begin{equation}
\begin{split}
\left\langle P^{\prime},S\left|T_{q, g,\text{Bel}}^{\mu \nu}\right| P,S\right\rangle=&2 A_{q, g}\left(0\right) \bar P^{\mu} \bar{P}^{\nu}\left(1+\epsilon^{0\rho\alpha\beta} \frac{i\Delta_\rho}{2M(\bar P^0+M)}S_\alpha \bar P_\beta\right) \\
&-\left(A_{q,g}(0)+B_{q, g}(0)\right)\frac{\bar P^{(\mu}}{M} \epsilon^{\nu)\alpha\beta\rho}i\Delta_\alpha S_\beta \bar P_\rho +O(\Delta^2) + g^{\mu\nu}\text{ terms}\ .
\end{split}
\end{equation}
Then the following result can be found with eq.~(\ref{jznormalized})
\begin{equation}
\begin{split}
\label{jzformfactors}
J^z_{q,g}=\bra{P,S}J^z_{q,g}\ket{P,S}=\frac{1}{2} \left(A_{q,g}(0)+B_{q,g}(0)\right)\ .
 \end{split}
\end{equation}
The AM derived this way is covariant, since the expansion is frame independent. Besides, the leading contribution to the AM in the infinite momentum frame which can be expressed in terms of parton densities as in the Jaffe-Manohar sum rule is also of interest. %

\subsection{Covariant twist-3 relations for quark contributions}
The symmetric EMT of the quark fields can be written as,
\begin{equation}
T^{\mu\nu}_{q,\text{Bel}}=\bar \psi \gamma^{(\mu} i \overleftrightarrow{D}^{\nu)}\psi\ ,
\end{equation}
where the covariant derivative was defined in section \ref{3fieldparam}.  The terms that contribute to the longitudinal AM are $T^{0y}_{q,\text{Bel}}$ and $T^{0x}_{q,\text{Bel}}$, which are equivalent. Without loss of generality, $T^{+y}_{q,\text{Bel}}$, the leading twist contribution to $T^{0y}_{q,\text{Bel}}$, will be discussed. The light-front correlations defined as
\begin{equation}
\begin{split}
\label{twist3emt}
T^{+y}_{q,LF,\text{Bel}}(x,y)=\int \frac{\text{d}\lambda \text{d}\mu}{(2\pi)^2}e^{i\frac {\lambda} {2} (x+y)+i\mu(y-x)}\bar\psi\left(-\frac{\lambda n}{2}\right) \gamma^{(+}W_{-\frac{\lambda}{2},\mu} i \overleftrightarrow{D}(\mu n)^{y)}W_{\mu,\frac{\lambda}{2}}  \psi\left(
\frac{\lambda n}{2}\right)\ ,
\end{split}
\end{equation}
can be introduced, which is defined as the correlation of three fields separated on the light front and the local EMT operator can be obtained by integrating over $x$ and $y$: $T^{+y}_{q,\text{Bel}}(0)=\int \text{d}x \text{d}y T^{+y}_{q,LF,\text{Bel}}(x,y)$. The $T^{+y}_{q,LF,\text{Bel}}(x,y)$ here does not have a simple parton picture since it involves three partons interacting with each other.

Unlike transverse polarization, there is no leading twist sum rule for longitudinal polarization since $J^z$ is invariant under boosts in the $z$ direction, whereas the transverse momentum grows as $J'^x\sim \gamma J^x$. Therefore, longitudinal polarization is intrinsically related to twist-3 GPDs. The 
definition in eq.~(\ref{twist3emt}) relates the EMT with the twist-3 GPDs defined in eq.~(\ref{qqDgpd}), and the following constraints from the gravitational form factors in eq.~(\ref{Tmunuformfactors}) follow
\begin{align}
\label{G3+perpsum}
  \int \text{d}x\text{d}y\; G_{q,D,3}(x,y)+\int \text{d}x \;x\; G_{q,3}(x)&= A_q(0)+B_q(0)\ ,\\
  \label{Gj+perpsum}
  \int \text{d}x\text{d}y G_{q,D,j}(x,y)=\int \text{d}x \;x\;G_{q,j}(x)&=  0\ ,
\end{align}
where $j=\{1,2,4\}$, which is consistent with the relations acquired from the discrete symmetry. The notation $G_{q,i}(x)$ is and will be used to express the forward limit of the GPD: $\lim_{\Delta \to 0} G_{q,i}(x,t,\xi)$. With the parameterization in eq.~(\ref{tensor+perpexp}), the EMT expressed with eq.~(\ref{twist3emt}) and the expression of AM eq.~(\ref{jznormalized}), $J_{q}^z$ can be derived as
\begin{equation}
\begin{split}
\label{jzsumrule1}
J_{q}^z=\frac{1}{2}\left(\int \text{d}x\text{d}y\; G_{q,D,3}(x,y)+\int \text{d}x \;x\; G_{q,3}(x)\right)= \frac{1}{2}(A_q(0)+B_q(0))\ .
\end{split}
\end{equation}
This result is consistent with the one derived with gravitational form factors and an equivalent result has been discussed and derived in ref. \cite{Hatta2012} and in ref. \cite{PENTTINEN2000,Kiptily2002}.

%
With the goal of decomposing the quark AM in order to derive a partonic sum rule, it will be helpful to decompose the quark spin contribution \cite{JaffeMan90}. Recall that the AM tensor can be expressed with the Belinfante EMT $T^{\mu\nu}_{q,\text{Bel}}$ as in eq.~(\ref{AMdensity}). Meanwhile, since the Belinfante EMT for QCD is simply the symmetrized canonical EMT
\begin{equation}
\begin{split}
T^{\mu\nu}_{\text{Bel}}= T^{(\mu\nu)}\ ,
 \end{split}
\end{equation}
it can alternatively be written as
\begin{equation}
\begin{split}
T^{\mu\nu}_{\text{Bel}}= T^{\mu\nu}-T^{[\mu\nu]}\ .
 \end{split}
\end{equation}
Then the AM density can be expressed with the canonical EMT as
\begin{equation}
\begin{split}
\label{smunurhodef}
    M^{\mu\nu\rho}(\xi)=&\xi^\nu \left(T^{\mu\rho}(\xi)-T^{[\mu\rho]}(\xi)\right) - \xi^\rho \left(T^{\mu\nu}-T^{[\mu\nu]}(\xi)\right)\ ,\\
    =&\xi^\nu T^{\mu\rho}(\xi) - \xi^\rho T^{\mu\nu}+S^{\mu\nu\rho}(\xi)\ ,
\end{split}
\end{equation}
with 
\begin{equation}
\begin{split}
\label{spindensity}
   S^{\mu\nu\rho}(\xi)=-\left(\xi^\nu T^{[\mu\rho]}(\xi) - \xi^\rho T^{[\mu\nu]}\right)\ .
\end{split}
\end{equation}
For the Dirac bilinears in quark EMT, following identities can be useful
\begin{align}
\label{spinident1}
   &\bar\psi(\xi)[\gamma^{\alpha} \overrightarrow{D}^{\beta}(\xi)-\gamma^{\beta} \overrightarrow{D}^{\alpha}(\xi)]\psi(\xi)= i\epsilon^{\alpha\beta\rho\sigma} \bar{\psi}(\xi) \gamma_\rho \gamma^5 \overrightarrow{D}_{\sigma}(\xi) \psi(\xi)\ ,\\
   \label{spinident2}
   &\bar\psi(\xi)[\gamma^{\alpha} \overleftarrow{D}^{\beta}(\xi)-\gamma^{\beta} \overleftarrow{D}^{\alpha}(\xi)]\psi(\xi)= -i\epsilon^{\alpha\beta\rho\sigma} \bar{\psi}(\xi) \gamma_\rho \gamma^5 \overleftarrow{D}_{\sigma}(\xi) \psi(\xi)\ ,
\end{align}
for which the equation of motion $\slashed D \psi(\xi)=0$ is used. With these relations it follows that
\begin{equation}
\begin{split}
\label{spinpolar}
   S_q^{\mu\nu\rho}(\xi)=-\frac{1}{2}\epsilon^{\mu\nu\rho\sigma}\bar\psi(\xi)\gamma_\sigma \gamma^5\psi(\xi)+\frac{1}{2}\partial_{\beta}\left(x^{[\nu}\epsilon^{\rho]\mu\beta\sigma}\bar\psi(\xi)\gamma_\sigma \gamma^5\psi(\xi)\right)\ ,
\end{split}
\end{equation}
where the second term is a total derivative term called a super potential whose forward matrix element is shown to vanish with proper wave packet treatment in ref. \cite{SHORE2000,Bakker2004}. Then the spin contribution can be derived with eq.~(\ref{spinpolar}) as
\begin{equation}
\begin{split}
\label{jaxialcharge}
J^z_{q,S}=-\frac{1}{2}\epsilon^{0xy\sigma}\int\text{d}^3\boldsymbol\xi\bra{P,S} \bar\psi(\boldsymbol \xi)\gamma_\sigma \gamma^5\psi(\boldsymbol \xi) \ket{P,S}\ .
\end{split}
\end{equation}
The matrix element of the axial current can be expressed as
\begin{equation}
\begin{split}
\bra{P,S} \bar\psi(0)\gamma_\sigma \gamma^5\psi(0) \ket{P,S}= 2 M S_{\sigma} \Delta q\ ,
\end{split}
\end{equation}
where the quantity $\Delta q$ is the singlet axial charge with all
quark flavors summed over. The axial charge $\Delta q$ can be related to the PDF defined in ref. \cite{JAFFEJI1992} with the axial current as
\begin{equation}
\begin{split}
\label{quarkspinpdf}
\int \frac{\mathrm{d} \lambda}{2 \pi}  \mathrm{e}^{i \lambda x}\left\langle P S\right|\bar{\psi}(0)& \gamma_{\mu} \gamma_{5} \psi(\lambda n)\left| P S\right\rangle\\
&\equiv 2\left[g_{1}(x) M p_{\mu}(S \cdot n)+Mg_{\mathrm{T}}(x) S_{\perp \mu}+M^3 g_{3}(x) n_{\mu}(S \cdot n) \right]\ ,
\end{split}
\end{equation}
where we have slightly rewritten it with our normalization condition $S^2=-1$. Then one has the relation (see \cite{Filippone2001} for more discussion on this)
\begin{equation}
\begin{split}
\label{deltaqdef}
\Delta q= \int \text{d}x g_{1}(x)=\int \text{d}x g_{T}(x)\ .
\end{split}
\end{equation}
Though only the quark helicity distribution $g_1(x)$ is needed for longitudinal spin contributions, both $g_1(x)$ and $g_T(x)$ are defined here which will be useful for transverse spin contributions in the next section. 

The spin contribution given by PDF would indicate certain constraint on the GPDs according to eq.~(\ref{spinident1}) and eq.~(\ref{spinident2}), which can be shown to be (see \cite{Leader2012} for treatment with form factors on this)
\begin{equation}
\begin{split}
\label{spinsumrule}
\int \text{d} x \;x G_{q,3}(x)-\int \text{d}x\text{d}y G_{q,D,3}(x,y)= \Delta q\ .
\end{split}
\end{equation}
The above relation can be used to express the AM with combinations of GPDs and a PDF. For instance if one eliminates $G_{q,D,3}(x,y)$ with eq.~(\ref{jzsumrule1}) and eq.~(\ref{spinsumrule}), it can be found that
\begin{equation}
\label{jzampdf}
 J^z_q=\int \text{d} x \left(\;x G_{q,3}(x)-\frac{1}{2} g_1(x) \right)\ ,
\end{equation}
which has already been derived in the literature \cite{Kiptily2002,PENTTINEN2000,Hatta2012}. Alternatively, one could instead eliminate the $G_{q,3}(x)$ to get 
\begin{equation}
\begin{split}
\label{OAMsumrule}
\int \text{d}x\text{d}y G_{q,D,3}(x,y)=J_q^z - \frac{1}{2}\Delta q = L^z_{q}\ .
\end{split}
\end{equation}
The above relation motivates the definition of a gauge-invariant covariant OAM density, 
\begin{equation}
\begin{split}
\label{OAMsumrule2}
 L^z_{q}(x)= \frac{1}{2}\int \text{d}y \left[G_{q,D,3}(x,y) +G_{q,D,3}(y,x)\right] \ ,
\end{split}
\end{equation}
which is an alternative definition compared with the ones
in ref. \cite{Ji2012}. However, its partonic interpretation
is not simple.

\subsection{Quark canonical OAM density}
\label{OAMquark}

Since the above covariant OAM density does not have a simple parton picture, further decomposition is needed to find one that does. With eq.~(\ref{smunurhodef}), the quark longitudinal AM $J^z_{q}$ can be expressed as
\begin{equation}
\begin{split}
\label{jqzrest}
    J^z_q=\int\text{d}^3 \boldsymbol \xi \left(\xi^x T_q^{0y}(\boldsymbol \xi) - \xi^y T_q^{0x}(\boldsymbol \xi)-\frac{1}{2}\epsilon^{0xyz}\bar\psi(\boldsymbol \xi)\gamma_z \gamma^5\psi(\boldsymbol \xi)\right)\ .
\end{split}
\end{equation}
The last term corresponds to the contribution from spin as discussed and it can be expressed with the light front correlation in eq.~(\ref{quarkspinpdf}) as $\Delta q_z= \int \text{d}x g_{1}(x)$. $g_{1}(x)$ has a clear partonic interpretation as the spin contribution of quarks with momentum $xP^+$ indicated by eq.~(\ref{quarkspinpdf}). 

In order to find a complete partonic picture for quark AM, it is better to remove the covariant derivative $D^{\nu}( \xi)$ in the first two terms of eq.~(\ref{jqzrest}). Therefore, the EMT of the quark can be further decomposed as
\begin{equation}
\begin{split}
\label{Tmunuparton}
T^{\mu\nu}_q(\xi)=\bar \psi(\xi) \gamma^\mu i \overleftrightarrow{D}^{\nu} \psi(\xi)= \bar \psi(\xi) \gamma^\mu i\overleftrightarrow{\partial} ^{\nu} \psi(\xi)- g \bar \psi(\xi) \gamma^\mu A^{\nu} \psi(\xi)\ .
\end{split}
\end{equation}
Then with following definitions of AM densities 
\begin{align}
M^{\mu\nu\rho}_{q,L}(\xi)&\equiv\xi^\nu T^{\mu\rho}_q(\xi)-\xi^\rho T^{\mu\nu}_q(\xi)\ ,\\
M^{\mu\nu\rho}_{q,l}(\xi)&\equiv\xi^\nu \bar \psi(\xi) \gamma^\mu i\overleftrightarrow{\partial} ^{\nu}  \psi(\xi)-(\nu\leftrightarrow\rho)\ ,\\
M^{\mu\nu\rho}_{q,\rm{pot}}(\xi)&\equiv\xi^\nu \bar \psi(\xi) \gamma^\mu ig A^{\nu} \psi(\xi)-(\nu\leftrightarrow\rho)\ ,
\end{align}
as well as their matrix elements in the longitudinally polarized nucleon, 
\begin{align}
\label{lqzdef}
L_q^z&\equiv{\int\text{d}^3\boldsymbol\xi\bra{P,S}M^{0xy}_{q,L}(\boldsymbol \xi)\ket{P,S}}\ ,\\
\label{lpqzdef}
l_q^z&\equiv{\int\text{d}^3\boldsymbol\xi\bra{P,S}M^{0xy}_{q,l}(\boldsymbol \xi)\ket{P,S}} \ ,\\
\label{lqpot}
l_{q,\rm pot}^z&\equiv{\int\text{d}^3\boldsymbol\xi\bra{P,S}M^{0xy}_{q,\rm pot}(\boldsymbol \xi)\ket{P,S}} \ ,
\end{align}
the canonical OAM can be written as
\begin{align}
\label{quarkoamdec}
l_q^z=L_q^z+l_{q,\rm pot}^z \ .
\end{align}
 Unlike $L_q^z$, the canonical OAM $l_q^z$ only involves two quark fields so it has a clear partonic interpretation, which can be also be seen through its connection with TMD’s and Wigner distributions \cite{Lorce2011,Lorce2012,Ji2012,HATTA20121}. However, in order to relate them, an extra term $l_{q,\rm pot}^z$ will be needed which has been discussed in for instance \cite{Wakamatsu:2010qj,Wakamatsu:2010cb,Wakamatsu:2011mb,Wakamatsu:2012ve}. As pointed out there, removing this term $l_{q,\rm pot}^z$ from quarks and attributing it to gluons will transform the covariant AM sum rule into the canonical Jaffe-Manohar sum rule.

Therefore, light-front canonical AM densities $l_q^z(x)$ can be introduced \cite{HAGLER1998,BASHINSKY1998,Harindranath1999}, 
\begin{equation}
\begin{split}
\label{lqzparton}
l_q^z(x)=\frac{1}{P^+}\int \frac{\mathrm{d} \lambda\text{d}^2\boldsymbol \xi^\perp}{2 \pi}  \mathrm{e}^{i \lambda x}\left\langle P ,S\right|\bar{\psi}(0,\boldsymbol \xi^\perp) \gamma^{+}(\xi^x i\partial^y-\xi^y i\partial^x)  \psi(\lambda n,\boldsymbol \xi^\perp)\left| P ,S\right\rangle\ ,
\end{split}
\end{equation}
with gauge links implicit between fields.  Though the moment of $\xi^\perp$ indicates that this forward matrix element can only be defined as the limit of an off-forward matrix element (see for instance Appendix \ref{defam}), this expression indeed has a simple partonic interpretation. Since $l_q^z(x)$ is the combination of OAM operators  $(\xi^x i\partial^y-\xi^y i\partial^x)$ and a parton density $\bar{\psi}(0) \gamma^{+} \psi(\lambda n)$, it can be considered as the OAM density of a parton with momentum $xP^+$. This is not the case for $l_{q,\rm pot}^z$ or $L_q^z$, since they have $A^{x/y}(\xi)$ or $D^{x/y}(\xi)$ which leads to a third parton field involved.

However, this decomposition breaks the gauge invariance, so neither the expression for $l_{q,\rm pot}^z$ nor $l_{q}^z$ is gauge invariant. If one defines a new type of GPD to measure $l_{q,\rm pot}^z$ as
\begin{equation}
\int \frac{\text{d}\lambda \text{d}\mu}{(2\pi)^2}e^{i\frac {\lambda} {2} (x+y)+i\mu(y-x)}\bra{P',S'}\bar\psi\left(-\frac{\lambda n}{2}\right) W_{-\lambda/2,\mu}\gamma^\alpha  A^{\beta}(\mu n) W_{\mu ,\lambda/2 }\psi\left(
\frac{\lambda n}{2}\right) \ket{P,S}\ ,
\end{equation}
it cannot be measurable since it is not gauge invariant. In order to get a  measurable $l_{q,\rm pot}^z$, a GPD equivalent to the one above while manifestly gauge invariant is needed.

Given that the light front gauge $A^+=0$ will always be used, the relation $A^{\alpha}=\frac{G^{+\alpha}}{\partial^+}$ can be very useful to express the gauge field in a manifestly gauge invariant way, for which one has
\begin{equation}
A^\mu(y)=A^\mu(x)+ \int_{x}^y \text{d}z G^{+\mu}(z)\ ,
\end{equation}
where the gauge links reduce to unity in the light front gauge. This leads to the following identity 
\begin{equation}
\begin{split}
\bar\psi\left(-\frac{\lambda n}{2}\right)\gamma^\alpha ig  A^{\beta}(\mu n) \psi\left(
\frac{\lambda n}{2}\right) =&  \bar\psi\left(-\frac{\lambda n}{2}\right) \gamma^\alpha  i g \int^{\mu}_{\nu} \text{d}t G^{+\beta}(t n) \psi\left(
\frac{\lambda n}{2}\right)\\
&+\bar\psi\left(-\frac{\lambda n}{2}\right) \gamma^\alpha  i g A^{\beta}(\nu n)\psi\left(
\frac{\lambda n}{2}\right)\  .
\end{split}
\end{equation}
While the LHS is the light front correlation corresponding to $l_{q,\rm{pot}}^z$, the off-forward matrix element of the RHS is related to the F-type GPDs defined in eq.~(\ref{Ftypedef}). The boundary term in the second line does not depend on $\mu$ and corresponds to a zero-mode contribution \cite{JI20200mode} which will be omitted. This relation indicates that $l_{q,\rm pot}^z$ can be expressed with the F-type GPDs. With 
the parameterization of the F-type GPDs given in eq.~(\ref{quarkFexp}), following result can be derived
\begin{equation}
l_{q,\rm pot}^z= \int \text{d}x\text{d}y \text{P}\frac{1}{y-x} G_{q,F,3}(x,y)\ ,
\end{equation}
whereas $L^z_q$ is already given by the D-type GPDs as eq.~(\ref{OAMsumrule}).

Then the quantity $l_q^z(x)$ which has a simple partonic interpretation is ready for discussion. With the relation in eq.~(\ref{Tmunuparton}) which is true for non-local operators as well, $l_q^z(x)$ can be expressed as a combination of D-type and F-type GPDs
\begin{equation}
\label{lgzparton}
l_{q}^z(x)= \int \text{d}y G_{q,D,3}(x,y)+\int\text{d}y \text{P}\frac{1}{y-x} G_{q,F,3}(x,y)\ ,
\end{equation}
and the total OAM is simply the integral over all the parton OAM densities
\begin{equation}
l_{q}^z= \int \text{d} x l_{q}^z(x)\ .
\end{equation}
Although the expression for $l_{q}^z$ might be affected by the light front zero mode as in ref. \cite{JI20200mode}. The eq.~(\ref{lgzparton}) was first derived in ref. \cite{Ji20132} and ref. \cite{Hatta2012} and as pointed out there, it shows that the the D-type and F-type twist-three GPDs are indeed related to each other \cite{EGUCHI2007,Zhou2010}. 
%
\subsection{Covariant twist-3 relations for gluon contributions}
The gluon contribution to longitudinal AM can be derived similarly with the eq.~(\ref{jznormalized}). Consider that the gluon EMT can be expressed as
\begin{equation}
T^{\mu\nu}_{g}(\xi)= 2 \text{Tr}\left\{-G^{\mu\rho}(\xi)G^{\nu}_{\;\;\rho}(\xi)+\frac{1}{4} g^{\mu\nu} G^{\rho\sigma}(\xi)G_{\rho\sigma}(\xi)\right\}\ .
\end{equation}
When longitudinal AM is concerned, only the matrix element of  $T^{+\perp}_g$ will contribute and the $g^{\mu\nu}$ term can be dropped. Consequently, the gluon EMT can be expressed by the following light front correlation
\begin{equation}
\label{gluontwist3EMT}
T^{+\perp}_{g,LF}(x)=-\int \frac{ \text{d}\lambda }{2\pi}e^{i \lambda x}2\text{Tr}\left\{G^{+\rho}\left(-\frac{\lambda n}{2}\right)W_{-\frac{\lambda}{2},\frac{\lambda}{2}}G^{\perp}_{\;\;\rho}\left(\frac{\lambda n}{2}\right)\right\}\ ,
\end{equation}
with $T^{+\perp}_g(0)=\int \text{d}x T^{+\perp}_{g,LF}(x)$. Again, this relation indicates certain constraints to the gluon GPDs defined in eq.~(\ref{gluonGPDdef}), 
\begin{align}
  \int \text{d}x \;x\; G_{g,3}(x)&= A_g(0)+B_g(0)\ ,\\
  \int \text{d}x \;x\;G_{g,j}(x)&=  0\ ,
\end{align}
with $j=\{1,2,4\}$.
Then a similar derivation with the gluon GPD parameterization in eq.~(\ref{gluonparameter2}) will lead to the following covariant twist-3 relation for gluon AM analogous to eq.~(\ref{jzsumrule1})
\begin{equation}
\begin{split}
\label{jgzsum}
J_{g}^z=\frac{1}{2} \int \text{d}x\;x\; G_{g,3}(x)= \frac{1}{2}(A_g(0)+B_g(0))\ .
\end{split}
\end{equation}
Unlike the quark AM, the gluon AM consists only of the 2-field correlation. However, this does not mean that it has a simple partonic interpretation, which can be shown after further decomposition of the gluon AM is done.
\subsection{Gluon spin and canonical OAM densities}
\label{gluonamdecomp}
To find a partonic interpretation for the gluon contribution, decomposition of the gluon AM into its orbital and spin contributions is needed and the expression \cite{JaffeMan90}
\begin{equation}
\begin{split}
\label{gluonamdecompose}
 M_g^{\mu\nu\lambda}(\xi)&=\left(\xi^{\nu} 2\text{Tr}\left\{ G^{\mu\alpha} G^{\lambda}_{\;\;\alpha}\right\}-(\nu\leftrightarrow\lambda)\right)\\
 &=2\text{Tr}\bigg{\{}G^{\mu \alpha}\left(\xi^{\lambda} \partial^{\nu}-\xi^{\nu} \partial^{\lambda}\right) A_{\alpha}+G^{\mu \lambda} A^{\nu}+G^{\nu \mu} A^{\lambda}\bigg{\}}\\
 &\qquad+g\bar\psi\gamma^{\mu}(\xi^\nu A^\lambda -\xi^\lambda A^\nu)\psi\ ,
\end{split}
\end{equation}
is needed. Obviously, the decomposition breaks the gauge symmetry explicitly. The first term corresponds to the OAM contribution and the second two terms correspond to a spin, so the following quantities are defined 
\begin{eqnarray}
\label{gldef}
 M_{g,l}^{\mu\nu\lambda}(\xi)&\equiv&2\text{Tr}\left\{ G^{\mu \alpha}\left(\xi^{\lambda} \partial^{\nu}-\xi^{\nu} \partial^{\lambda}\right) A_{\alpha}\right\}\ ,\\
 \label{gsdef}
 M_{g,S}^{\mu\nu\lambda}(\xi)&\equiv&2\text{Tr}\left\{ G^{\mu \lambda} A^{\nu}+G^{\nu \mu} A^{\lambda}\right\}\ ,\\
 \label{glpotdef}
 M_{q,\rm{pot}}^{\mu\nu\lambda}(\xi)&\equiv& g\bar\psi\gamma^{\mu}(\xi^\nu A^\lambda -\xi^\lambda A^\nu)\psi\ .
\end{eqnarray}
Though each of the terms here is not gauge invariant, the total contribution is \cite{Hoodbhoy1999}. The $M^{\mu\nu\rho}_{q,\rm{pot}}$ term cancels the $l_{q,\rm{pot}}$ term that is artificially added in the quark AM decomposition subsection in eq.~(\ref{quarkoamdec}).

First consider the gluon spin term $M^{\mu\nu\rho}_S$, and it has been pointed out that it does not correspond to the U(1) axial current $K_{\sigma}$ in ref. \cite{JaffeMan90}. However, in the light cone gauge $A^{+}=0$ at leading twist where $\mu=+$, they are actually the same. Since the gluon axial current $K_{\sigma}$ can be written as
\begin{equation}
\begin{split}
\epsilon^{\mu \nu \lambda \sigma} K_{\sigma}\equiv2\text{Tr}\left\{G^{\mu \lambda} A^{\nu}+G^{\nu \mu} A^{\lambda}+G^{\lambda \nu} A^{\mu}\right\}\ ,
\end{split}
\end{equation}
its only difference from $M^{\mu\nu\rho}_{q,\rm{pot}}$ is the third term $G^{\lambda \nu} A^{\mu}$ which vanishes at leading twist with $\mu=+$ and in the light cone gauge $A^+=0$. Furthermore, this relation can be used to express the gluon spin contribution with the specific PDF defined by the operators above. 

Motivated by that, consider the polarized gluon distribution $\Delta G(x)$ defined as \cite{Manohar1990,JAFFE1996}
\begin{equation}
\begin{split}
\label{gluonhelicity}
\Delta G(x)=\frac{i}{ x P^+} \int \frac{\mathrm{d} \lambda}{ 2\pi} e^{i x \lambda}\bra{ P,S}2\text{Tr}\left\{ G^{+\alpha}\left(-\frac{\lambda n}{2}\right)W_{-\frac{\lambda}{2},\frac{\lambda}{2}} \tilde{G}_{\;\;\alpha}^{+}\left(\frac{\lambda n}{2}\right)\right\} \ket{P,S}\ ,
\end{split}
\end{equation}
where $\tilde G^{\mu\nu\alpha\beta}=\frac{1}{2}\epsilon^{\mu\nu\alpha\beta}G_{\alpha\beta}$ and as discussed in ref. \cite{Ji2013} the gluon helicity distribution $\Delta G(x)$ is both measurable and calculable on lattice. Then the matrix element of $M^{+xy}_{g,S}$ can be expressed with the above PDF and the relation \cite{LEADER20142}
\begin{equation}
J^z_{g,S}=\int \text{d}^3 \boldsymbol\xi\bra{P,S}M^{+xy}_{S}(\boldsymbol \xi)\ket{P,S} =\Delta G\equiv \int \text{d} x\Delta G(x)\ ,\end{equation}
can be derived. This expression, associated with eq.~(\ref{gluonhelicity}), gives a simple parton picture for the gluon helicity contribution, with $\Delta G(x)$ the contribution from gluon spin of partons with momentum $x P^+$. Similarly to the quark case, a partonic interpretation for gluon OAM is need then.
%

Again, the gluon OAM can be expressed with the gluon OAM density as
\begin{equation}
l^{z}_g=\bra{P,S}\int \text{d}^3 \boldsymbol \xi M^{0xy}_{g,l}(\boldsymbol \xi)\ket{P,S}
=\bra{P,S}\int \text{d}^3 \boldsymbol\xi 2\text{Tr}\left\{G^{0 \alpha}\left(\xi^{y} \partial^{x}-\xi^{x} \partial^{y}\right) A_{\alpha}\right\}\ket{P,S}\ ,
\end{equation}
which can be expressed with a light front correlation defined as
\begin{equation}
l_g^z(x)=\int \frac{\mathrm{d} \lambda\text{d}^2\boldsymbol \xi^\perp}{2 \pi}  \mathrm{e}^{i \lambda x}\left\langle P, S\right|2\text{Tr}\left\{G^{\;\;+\alpha}(0,\boldsymbol \xi^\perp) (\xi^x \partial^y-\xi^y \partial^x)  A_{\alpha}(\lambda n,\boldsymbol \xi^\perp)\right\}\left| P ,S\right\rangle\ ,
\end{equation}
while $l^{z}_g$ is simply the integral of $l^{z}_g(x)$ over $x$. The  expression of $l_g^z(x)$ has a simple partonic interpretation, since it is just the OAM operator $(\xi^x \partial^y-\xi^y \partial^x)$, combining the gluon density $2\text{Tr}\left\{G^{+\alpha}(0)A_{\alpha}(\lambda n)\right\}$ of momentum $xP^+$.

However, the $l^{z}_g(x)$ defined here is obviously not gauge invariant. In order to express $l^{z}_g(x)$ with gauge invariant quantities, consider the following relation in the light front gauge
\begin{equation}
\label{giegluon}
G^{\mu \alpha} i\partial^{\nu}  A_{\alpha}=G^{\mu \alpha} i D^{\nu} \frac{1}{\partial^+}G^{+}_{\;\;\alpha}+ G^{\mu \alpha} g\frac{1}{\partial^+} G^{+\nu} \frac{1}{\partial^+}G^{+}_{\;\;\alpha} \ ,
\end{equation}
and its non-local version
\begin{equation}
\begin{split}
\label{GIEgluonOAM}
G^{\alpha \sigma}\left(-\frac{\lambda}{2} n\right) i\partial^{\beta}  A_{\sigma}\left(-\frac{\lambda}{2} n\right)=&G^{\alpha \sigma}\left(-\frac{\lambda}{2} n\right) iD^{\beta}(\mu n) \frac{1}{\partial^+}G^{+}_{\;\;\sigma}\left(\frac{\lambda}{2} n\right)\\
&+G^{\alpha \sigma}\left(-\frac{\lambda}{2} n\right) g \frac{1}{\partial^+}G^{+\beta}(\mu n) \frac{1}{\partial^+}G^{+}_{\;\;\sigma}\left(\frac{\lambda}{2} n\right)\ .
\end{split}
\end{equation}
The operators on the RHS correspond to the D/F-type gluon GPDs defined in eq.~(\ref{gluonDtypedef}) and eq.~(\ref{gluonFtypedef}) again with zero mode contribution omitted, while the operator on the LHS gives us the gluon OAM operator in the local limit. Therefore, this relation can be used to measure gluon OAM with measurable D-type and F-type gluon GPDs. In order for that, two auxiliary quantities can be defined as
\begin{align}
M^{\mu\nu\rho}_{g,\tilde L}(\xi)&\equiv2\text{Tr}\left\{ G^{\mu \alpha}\left(\xi^{\lambda} D^{\nu}-\xi^{\nu} D^{\lambda}\right) A_{\alpha}\right\}\ ,\\
M^{\mu\nu\rho}_{g,\rm{pot}}(\xi)&\equiv -2ig\text{Tr}\left\{G^{\mu \alpha} \left(\xi^{\lambda}  A^{\nu}-\xi^{\nu}A^{\lambda}\right) A_{\alpha}\right\}\ ,
\end{align}
with their matrix element defined as
\begin{align}
\label{lgzdef}
\tilde L_g^z&\equiv{\int\text{d}^3\boldsymbol\xi\bra{P,S}M^{0xy}_{g,\tilde L}(\boldsymbol \xi)\ket{P,S}}\ ,\\
\label{lgzpotdef}
l_{g,\rm{pot}}^z&\equiv{\int\text{d}^3\boldsymbol\xi\bra{P,S}M^{0xy}_{g,\rm{pot}}(\boldsymbol \xi)\ket{P,S}}\ .
\end{align}
According to the relation in eq.~(\ref{GIEgluonOAM}), $\tilde L_g^z(x)$ and $l_{g,pot}^z(x)$ can be expressed with the D-type and F-type gluon GPDs as
\begin{align}
\tilde L_g^z&(x)= \int\text{d}y\text{P}\frac{1}{x+y} G_{g,D,3}(x,y),\\
l_{g,\rm{pot}}^z&(x)= \int\text{d}y \text{P}\frac{1}{y^2-x^2}G_{g,F,3}(x,y)\ ,
\end{align}
which will finally gives the expression of $l_g^z(x)$ in terms of the D-type and F-type GPDs as
\begin{align}
 l_g^z(x)=\int \text{d}y  \text{P}\frac{1}{x+y} G_{g,D,3}(x,y)+\int \text{d}y  \text{P}\frac{1}{y^2-x^2}G_{g,F,3}(x,y)\ ,
\end{align}
with the total gluon OAM $l_g^z$ as an integral of gluon OAM density $l_g^ z(x)$:
\begin{align}
l_g^z=\int \text{d}x l_g^z(x)\ .
\end{align}
Though again, the integral might suffer from the zero mode problem just like the quark case \cite{JI20200mode}. Equivalent expressions are first derived in ref. \cite{Ji20132}  and ref. \cite{Hatta2012} (with a different notation of $L_{can}^g$).  
\section{A twist-three spin sum rule for transversely-polarized nucleon}
\label{sec4}

Spin sum rules for a transversely polarized nucleon have received 
much less attention in the community, and some of their theoretical results have been hotly 
debated over the years \cite{Bakker2004,Burkardt2003,Burkardt2005,Leader2012,Ji2012,Ji:2012vj,Leader:2012md,Hatta2013,Ji2020}.  It turns out that one of the key issues in this case
is the separation of the intrinsic spin from the center-of-mass 
contributions when the nucleon is in motion. [For the nucleon at zero momentum, 
there is no distinction between transverse and longitudinal polarization.] 
For a transversely-polarized nucleon, its overall ``center-of-mass motion" will
generate a contribution to the AM as well as to any quantity that depends on
the polarization. Fortunately, Lorentz symmetry allows the intrinsic and extrinsic 
contributions to be distinguished through their dependencies on the different 
Lorentz structures. As explained in ref. \cite{Ji2020}, the Pauli-Lubanski vector (\ref{PVvector})
was introduced to exclude extrinsic sources of AM automatically. 
The leading-twist partonic sum rule for the transverse AM in the infinite momentum 
frame has been re-derived in terms of this strategy.  
In this section, we will give a detailed argument for this and also 
derive a new twist-3 spin sum rule for the transverse spin. 

The transverse AM contains both twist-2 and twist-3 contributions. To see this, 
one can write the forward matrix element of transverse AM as (See Appendix \ref{defam} for discussion and derivation) %
\begin{equation}
\begin{split}
\label{jxqnormalized}
J^x_{q,g}\equiv\bra{P,S}J^x_{q,g}\ket{P,S}=\frac{1}{2E_P}&\left(\lim_{\Delta\to 0,S'\to S}i\frac{\partial}{\partial \Delta_y}\bra{P+\Delta,S'} T_{q,g,\text{Bel}}^{0z}(0)  \ket{P,S}\right. \\
&\left.\;\;\;\;-\lim_{\Delta\to 0,S'\to S}i\frac{\partial}{\partial \Delta_z}\bra{P+\Delta,S'} T_{q,g,\text{Bel}}^{0y}(0)  \ket{P,S}\right)\ ,
 \end{split}
\end{equation}
where we assumed the transverse polarization is in the $x$ direction while the other transverse direction is $y$. The fact that the two terms here corresponding to GPDs of different twist is made more clear
in light front coordinates where the substitutions $0z\rightarrow++$ and $0y\rightarrow+\perp$ can 
be made. Defining the first and second terms with their twist number superscripts explicit, then
\begin{equation}
J_{q,g}^x=J_{q,g}^{x(2)}+J_{q,g}^{x(3)}\ ,
\end{equation}
where
\begin{align}
    \label{j2xdef}
    &J_{q,g}^{x(2)}\equiv\bra{P,S}\hat{J}_{q,g}^{x(2)}\ket{P,S}=\frac{1}{2E_P}\lim_{\Delta\to 0,S'\to S}i\frac{\partial}{\partial \Delta_y}\bra{P+\Delta,S'} T_{q,g,\text{Bel}}^{0z}(0)  \ket{P,S}\ ,\\
    \label{j3xdef}
    &J_{q,g}^{x(3)}\equiv\bra{P,S}\hat{J}_{q,g}^{x(3)}\ket{P,S}=-\frac{1}{2E_P}\lim_{\Delta\to 0,S'\to S}i\frac{\partial}{\partial \Delta_z}\bra{P+\Delta,S'} T_{q,g,\text{Bel}}^{0y}(0)  \ket{P,S}\ .
\end{align}
Again the twist counting is made by the components of the EMT in the infinite momentum frame. 

In the rest frame the system has rotational symmetry, so the two terms which are related by the exchange of indices should have the same contribution
\begin{equation}
J_{q,g}^{x(2)}(P^z=0)=J_{q,g}^{x(3)}(P^z=0)\ ,
\end{equation}
and as a result $J^x_{q,g}(P^z=0)=2J_{q,g}^{x(2)}(P^z=0)=2J_{q,g}^{x(3)}(P^z=0)$. As we shall see, 
these two quantities can lead to two separate transverse spin sum rules in the infinite momentum frame.

\subsection{Intrinsic transverse AM}
\label{intrinsic}
The transverse AM can be derived from gravitational form factors the same way as the longitudinal case, 
and the result for the twist-2 term is
\begin{equation}
\begin{split}
\label{jx2ff}
J^{x(2)}_{q,g}=\frac{{P^0}^2+{P^z}^2}{4MP^0} \left(A_{q,g}(0)+B_{q,g}(0)\right)-\frac{{P^z}^2}{2M(P^0+M)}A_{q,g}(0)\ ,
 \end{split}
\end{equation}
and for the twist-3 term, 
\begin{equation}
\begin{split}
\label{jx3ff}
J^{x(3)}_{q,g}=\frac{{P^0}^2-{P^z}^2}{4MP^0} \left(A_{q,g}(0)+B_{q,g}(0)\right)\ .
 \end{split}
\end{equation}
The $(P^z)^2$ term in the twist-3 case needs a bit of explanation. With $\vec{P}'$
in the $z$ direction, $\Delta^0$ cannot vanish. Rather it is constrained to be related to 
$\Delta^z$,
\begin{equation}
            \Delta^0 \bar{P}^0 - \Delta^z \bar{P}^z =0 \ .
\end{equation}
Therefore, the derivative with respect to $\Delta^z$ here will generate a contribution 
proportional to $P^z$. 

The total transverse AM can be expressed as
\begin{equation}
\begin{split}
J^x_{q,g}=\frac{P^0}{2M} \left(A_{q,g}(0)+B_{q,g}(0)\right)-\frac{{P^z}^2}{2M(P^0+M)}A_{q,g}(0)\ .
 \end{split}
\end{equation}
This is consistent with the result in ref. \cite{Leader2014}, where an extra term proportional to $P^z$ is present in the expression. However, this contribution comes from the center-of-mass motion, and shall be 
removed when discussing the intrinsic spin sum rule \cite{Ji2020}. 
Motivated by the Pauli-Lubanski vector in eq.~(\ref{PVvector}), one can simply drop
terms proportional in $P^{i}$ or $P^{j}$ in $J^{ij}$. In the case of transverse AM in the $x$ direction, one has $i=y$ and $j=z$, so terms proportional to $P^z$ are dropped. It is also clear if one writes 
the matrix element of AM covariantly as \cite{Leader2014}
\begin{equation}
\begin{split}
J^{ij}_{q,g}=-\frac{\epsilon^{ij\alpha\beta}}{2M} S_\alpha P_\beta \left(A_{q,g}(0)+B_{q,g}(0)\right)-\frac{\boldsymbol {P}^i(\boldsymbol P\times \boldsymbol s)^j-\boldsymbol {P}^j(\boldsymbol P\times \boldsymbol s)^i}{2M(P^0+M)}A_{q,g}(0)\ ,
 \end{split}
\end{equation}
where the last term is apparently proportional to $P^{i}$ or $P^{j}$ and is thus non-intrinsic. After removing the non-intrinsic contribution, the intrinsic transverse AM will grow with a factor of $\gamma$ as the nucleon gets boosted, as discussed in ref. \cite{Ji2020}.

However, the process of taking the intrinsic part actually interferes with the twist counting. 
Besides the term one eventually removes in eq.~(\ref{jx2ff}), there are other terms proportional to $P^{z}$ in eq.~(\ref{jx2ff}) and eq.~(\ref{jx3ff}) which happen to cancel in the end. One can explicitly check 
that these terms arise from contributions to $J^{\mu\nu}$ through a $P^\mu$ or $P^\nu$ type 
of dependence. If one retains those terms as intermediate results and allows them to cancel 
in the end, the two contributions will be
\begin{align}
\label{jxqtw2}
J^{x(2)}_{q,g}&=\left(\frac{\gamma}{2}-\frac{1}{4\gamma}\right) \left(A_{q,g}(0)+B_{q,g}(0)\right)\ ,\\ J^{x(3)}_{q,g}&=\frac{1}{4\gamma} \left(A_{q,g}(0)+B_{q,g}(0)\right) \ .
\end{align}
 Only the leading term contributes in the infinite momentum limit, while the other term corresponds to 
 a suppressed contribution in terms of large-momentum power counting. However,
 the precise $1/\gamma$ dependence in the above equations guarantees the total AM $J^{x}_{q,g}$
 to have a simple Lorentz transformation property as predicted by Lorentz symmetry~\cite{Ji2020}.
 Therefore, the leading and higher order terms must separately give the same AM to maintain
 Lorentz symmetry, and the two terms shall lead to two different transverse AM sum rules 
 respectively.

One could also remove the non-intrinsic contribution from each individual term,
 \newline eq.~(\ref{jx2ff}) and eq.~(\ref{jx3ff}) will then correspond to 
\begin{align}
J^{x(2)}_{q,g,\rm{In}}&=\frac{\gamma}{4}\left(A_{q,g}(0)+B_{q,g}(0)\right)\ ,\\ J^{x(3)}_{q,g,\rm{In}}&=\frac{\gamma}{4} \left(A_{q,g}(0)+B_{q,g}(0)\right) \ ,
\end{align}
where the subscript ``$\rm{In}$" indicates taking the intrinsic part of each term respectively. Then the twist-3 term is promoted to contribute at leading twist and it has the same contribution as the twist-2 part just as in the rest frame case. Still, the total transverse AM has a simple $\gamma$ dependence, but in this picture both contributions are needed in order to have a complete transverse AM even in the infinite momentum frame.

To summarize, if one removes the center-of-mass contribution in the end, 
the transverse AM will be dominated by the twist-2 contribution. Then a 
simple parton picture for transverse polarization emerges from the twist-2 
GPDs~\cite{Burkardt2005,Ji2012,Ji:2012vj}, as we shall explain in the next 
subsection. On the other hand, Lorentz symmetry also allows us
to derive a twist-3 parton transverse spin sum rule  
as we demonstrate in the other two subsections, which is the analogue of  
the Jaffe-Manohar spin sum rule for the longitudinal polarization.
On the other hand, if taking the intrinsic part of the two terms respectively, 
one ends up with a transverse AM sum rule with one half of the contribution from 
the twist-2 and another one half from the twist-3~\cite{Ji2020}, which is nothing but the average of
two spin sum rules in the previous case.

\subsection{Transverse AM parton density and twist-2 sum rule}
\label{twist2sumrule}
The twist-2 partonic contribution to transverse AM has already been well-established \cite{Ji2012,Ji:2012vj, Ji2020}. Here we revisit it to emphasize the procedure to isolate the intrinsic contribution. The covariant sum rule derived in ref. \cite{Ji1997} is gauge invariant 
and frame independent, but it does not provide a simple parton picture. The 
lesson from the Jaffe-Manohar sum rule for longitudinal polarization is that 
the canonical AM can be easily related to partons~\cite{JaffeMan90}.  
However, for the transverse polarization, covariant sum rule  
in the infinite momentum frame does provide a simple parton interpretation~\cite{Burkardt2005, Ji2012}. 

Consider the twist-2 $J^{x(2)}_{q,g}$ defined in eq.~(\ref{j2xdef}), in which the term $T_{q,g,\text{Bel}}^{0z}(0)$ is involved. In the infinite momentum frame, one can generate this term from a light-front correlation defined as
\begin{equation}
\begin{split}
 T_{q,LF,\text{Bel}}^{0z}(x,y)  &=\int \frac{ \text{d}\lambda \text{d}\mu}{(2\pi)^2}e^{i\frac {\lambda} {2} (x+y)+i\mu(y-x)}\bar\psi\left(-\frac{\lambda n}{2}\right) W_{-\frac{\lambda}{2},\mu}\gamma^{(0} i\overleftrightarrow{D}^{z)}(\mu n) W_{\mu,\frac{\lambda}{2}}\psi\left(
\frac{\lambda n}{2}\right)\ ,
\end{split}
\end{equation}
while $T^{0z}_{q,\text{Bel}}=\int\text{d}x\text{d}y T_{q,LF,\text{Bel}}^{0z}(x,y)$. Now consider the leading twist contribution to $T^{0z}_{q,\text{Bel}}$ for which $T_{q,LF,\text{Bel}}^{++}(x)$ will be concerned and one has 
\begin{equation}
\begin{split}
 T_{q,LF,\text{Bel}}^{++}(x)  =\int \frac{ \text{d}\lambda}{2\pi}e^{i \lambda  x}\bar\psi\left(-\frac{\lambda n}{2}\right) \gamma^+ i\overleftrightarrow{\partial}^+ \psi\left(\frac{\lambda n}{2}\right)\ ,
\end{split}
\end{equation}
where the light front gauge $A^+=0$ is chosen, so the covariant derivative $\overleftrightarrow{D}^+(\mu n)$ reduces to a partial derivative $\overleftrightarrow{\partial}^+$ and the gauge links reduces to unity. Then the leading twist EMT can be expressed as $T^{++}_{q,\text{Bel}}=\int\text{d}x T_{q,LF,\text{Bel}}^{++}(x)$, which has a simple parton picture with $i\partial^+$ the momentum operator combining with a quark density operator of momentum $xP^+$. The above expression leads to a definition of 
the parton AM density, 
\begin{equation}
\begin{split}
\label{jx2lf}
 J^{x(2)}_{q}(x)=\int \frac{ \text{d}\lambda\text{d}^2\boldsymbol \xi^\perp}{2\pi}e^{i \lambda  x}\bra{P,S}\bar\psi\left(-\frac{\lambda n}{2},\boldsymbol \xi^\perp\right) \gamma^+ \xi^y i\overleftrightarrow{\partial}^+ \psi\left(\frac{\lambda n}{2},\boldsymbol \xi^\perp\right)\ket{P,S}\ ,
\end{split}
\end{equation}
with the total transverse AM given by $J^{x(2)}_{q}=\int{\text{d}}x J^{x(2)}_{q}(x)$. Thus, the covariant sum rule has a simple parton picture without decomposing the AM into spin and orbital components.
The RHS of eq.~(\ref{jx2lf}) involves a moment of $\xi$ and it can only be defined as the limit of an off-forward matrix element. Thus in order to derive $J^{x(2)}_{q}(x)$, one needs to work with the off-forward matrix element of the RHS, which gives the well-known twist-2 GPDs. With eq.~(\ref{gamma+exp}) and eq.~(\ref{gluonparameter1}), the twist-2 quark GPDs to the leading order of $\Delta$ can be written as
\begin{equation}
\begin{split}
\label{twist2exp}
&\int_{-\infty}^{\infty}\frac{\text{d}\lambda}{2\pi}e^{i\lambda x}\bra{P',S'}\bar\psi\left(-\frac{\lambda n}{2}\right)\gamma^+ \psi\left(
\frac{\lambda n}{2}\right) \ket{P,S}\\
&=   \frac{\bar P^+}{M}H_q(x) \left(2M+\epsilon^{0\rho\alpha\beta} \frac{i\Delta_\rho}{\bar P^0+M}S_\alpha \bar P_\beta\right)-\left(H_q(x)+E_q(x)\right) i \frac{\epsilon^{+\nu\alpha\beta}}{M}\Delta_\nu S_\alpha \bar P_\beta \ ,
\end{split}
\end{equation}
which combined with eq.~(\ref{jx2lf}) gives
\begin{equation}
\begin{split}
\label{jx2qlf}
 J^{x(2)}_{q}(x)=\gamma \frac{x}{2} \left(H_{q}(x)+E_{q}(x)\right)\ ,
 \end{split}
\end{equation}
where the intrinsic contribution of AM is isolated by removing the term proportional to $P^{\mu}$ in $J^{\mu\nu}$ as discussed in the previous subsection. The gluon twist-2 AM can be derived in the same way~\cite{Ji1997} and the result looks similar. 
$J^{x(2)}_{q,g}(x)$ clearly represents the twist-2 contribution to transverse AM density
from partons with momentum $xP^+$ while the extra $\gamma$ factor corresponds to the boost effect it possesses. If one compares the GPD parameterization in eq.~(\ref{gamma+exp}) and  eq.~(\ref{gluonparameter1}) with the gravitational form factors in eq.~(\ref{Tmunuformfactors}), one acquires the constraints \cite{Ji1997}: $
  \int \text{d}x \;x H_{q,g}(x)=  A_{q,g}(0) $ and $
   \int \text{d}x \;x E_{q,g}(x)=  B_{q,g}(0)$. 
These constraints can be used to express the twist-2 contribution to transverse AM with gravitational form factors, producing eq.~(\ref{jxqtw2}).
\subsection{Quark twist-3 transverse-AM densities}
 It is straightforward to derive the twist-3 contribution $J_{q,g}^{x(3)}$ directly from eq.~(\ref{j3xdef}) with $T^{0y}_{q,\text{Bel}}$ the same as for the longitudinal AM in eq.~(\ref{twist3emt}). The covariant relation for the transverse AM is 
\begin{equation}
\begin{split}
\label{jxtwist3}
J_{q}^{x(3)}=\frac{1}{4\gamma}\left[\int \text{d}x\text{d}y\; G_{q,D,3}(x,y)+\int \text{d}x \;x\; G_{q,3}(x)\right]\ ,
\end{split}
\end{equation}
analogous to eq.~(\ref{jzsumrule1}) in the longitudinal case. According to the discussion in section \ref{intrinsic}, these twist-3 expressions are always associated with a factor of $\frac{1}{4\gamma}$ or $\frac{\gamma}{4}$ depending on how one deals with the center-of-mass motion. When the twist-3 term is kept as a higher twist contribution, the quantities are associated with a $\frac{1}{4\gamma}$ factor. We shall suppress the $\frac{1}{2\gamma}$ factor in the future discussion,
so that it is commensurate with the longitudinal sum rule in eq.~(\ref{jzsumrule1}) after the renormalization.

To find a simple parton picture, one needs to go through the same AM decomposition process in canonical quantities as in the case of longitudinal AM \cite{JaffeMan90}, expecting a new sum rule for the transverse AM in terms of twist-3 quantities,    
\begin{align}
\label{jmsumrulex}
\frac{1}{2}\Delta q_T +\Delta G_T +  +l^{x(3)}_q+l^{x(3)}_g =\frac{\hbar}{2} \ ,
\end{align}
with each term practically the same as that in eq.~(\ref{jmsumrule}) because of the rotational symmetry, but has a different partonic interpretation due to dynamical boost. 

First, consider the twist-3 transverse AM of quarks, which can be written as
\begin{equation}
\begin{split}
\label{jx3parton2}
\hat  J^{x(3)}_{q}=\int \text{d}^3\boldsymbol \xi \left[-\bar \psi(\boldsymbol\xi)\gamma^0\left( \xi^z \times i\overleftrightarrow{D}^y(\boldsymbol\xi)\right) \psi(\boldsymbol\xi)+\frac{1}{4}\epsilon^{0yzx}\bar \psi(\boldsymbol\xi)\gamma_x \gamma_5   \psi(\boldsymbol\xi) \right]\ ,
\end{split}
\end{equation}
which, similar to eq.~(\ref{jqzrest}), allows for further decomposition in canonical terms. Defining matrix elements 
similar to eq.~(\ref{lqzdef})-eq.~(\ref{lqpot})
\begin{align}
l_q^{x(3)}&\equiv-\int\text{d}^3\boldsymbol\xi\bra{P,S}\bar \psi(\boldsymbol\xi)\gamma^0\left( \xi^z \times i\overleftrightarrow{\partial}^y\right) \psi(\boldsymbol\xi)\ket{P,S}\ ,\\
L_q^{x(3)}&\equiv- \int\text{d}^3\boldsymbol\xi\bra{P,S}\bar \psi(\boldsymbol\xi)\gamma^0\left( \xi^z \times i\overleftrightarrow{D}^y(\boldsymbol\xi)\right) \psi(\boldsymbol\xi)\ket{P,S}\ ,\\
l_{q,\rm{pot}}^{x(3)}&\equiv-\int\text{d}^3\boldsymbol\xi\bra{P,S}\bar \psi(\boldsymbol \xi)\gamma^0\left( \xi^z \times g A^y(\boldsymbol \xi)\right) \psi(\boldsymbol\xi)\ket{P,S}\ .
\end{align}
Clearly $l_q^{x(3)}=L_q^{x(3)}+l_{q,\rm{pot}}^{x(3)}$ according to the definition of the covariant derivative. Then the twist-3 AM $J^{x(3)}_{q}$can be expressed with the quantities defined above as
\begin{equation}
\begin{split}
\label{jqx3decomp}
 J^{x(3)}_{q}= l_q^{x(3)}-l_{q,\rm{pot}}^{x(3)}+\frac{1}{2}\Delta q_T \ .
\end{split}
\end{equation}
The $l_{q,\rm{pot}}^{x(3)}$ term will cancel the term that shows up when decomposing the gluon AM which will be discussed then. As shown in eq.~(\ref{quarkspinpdf}), the spin contribution $\Delta q_T=\Delta q $ can be expressed with $g_T(x)$ which represents the spin contribution from quarks with momentum $x P^+$.

As for $l^{x(3)}_{q}$, it can be expressed in terms of the parton density defined as 
\begin{equation}
\begin{split}
\label{lqx3parton}
l_q^{x(3)}(x)=-\frac{1}{P^+}\int \frac{\mathrm{d} \lambda\text{d}^2\boldsymbol \xi^\perp}{2 \pi}  \mathrm{e}^{i \lambda x}\left\langle P S\right|\bar{\psi}(0,\boldsymbol \xi^\perp) \gamma^{+}\xi^z i\partial^y \psi(\lambda n,\boldsymbol \xi^\perp)\left| P S\right\rangle\ ,
\end{split}
\end{equation}
which is similar to eq.~(\ref{lqzparton}), and $l^{x(3)}_{q}=\int \text{d}x l^{x(3)}_{q}(x)$. This density has a simple parton picture with $\xi^z i\partial^y$ the twist-3 part of the transverse AM operator and momentum fraction $xP^+$. The two auxiliary quantities $L_{q}^{x(3)}$ and $l_{q,\rm{pot}}^{x(3)}$ are related to three-parton correlation functions.

The expression of $l_q^{x(3)}(x)$ in eq.~(\ref{lqx3parton}) 
can be calculated using the twist-3 GPDs,  
\begin{equation}
    \begin{split}
        l_{q}^{x(3)}(x)= \int \text{d}y\bigg{[}&\left(G_{q,D,3}(x,y)+G'_{q,D,4}(x,y)\right)
        \\&+ \text{P}\frac{1}{y-x} \left(G_{q,F,3}(x,y)+G'_{q,F,4}(x,y)\right)\bigg{]}\ ,
    \end{split}
\end{equation}
where $G'_{q,D/F,4}(x,y)\equiv \lim_{\Delta\to 0}\frac{\text{d}}{\text{d}\xi}G_{q,D/F,4}(x,y,t,\xi)$. 
This result is consistent with 
\begin{align}
L_{q}^{x(3)}&= \int \text{d}x\text{d}y \left(G_{q,D,3}(x,y)+G'_{q,D,4}(x,y)\right)\ ,\\
l_{q,\rm{pot}}^{x(3)}&= \int \text{d}x\text{d}y \text{P}\frac{1}{y-x} \left(G_{q,F,3}(x,y)+G'_{q,F,4}(x,y)\right)\ .
\end{align}
Though the integrals of $G'_{q,D/F,4}(x,y)$ vanish after integration over both $x$ and $y$, they reflect the fact that the transverse and longitudinal AM have different partonic densities. Once integrated over all $x$, $l_{q}^{x(3)}=l_{q}^{z}$ as required by
rotational symmetry.

%
%
%
\subsection{Gluon twist-3 transverse AM densities and  twist-3 spin sum rule}
In order to depict a full twist-3 parton picture of transverse AM, the contribution of gluons will be needed. By definition, the twist-3 gluon transverse AM can be written as
\begin{equation}
\label{jx3g}
\hat J^{x(3)}_{g}=\int \text{d}^3 \boldsymbol\xi \; 2\xi^{z} \text{Tr}\left\{  G^{0\alpha}(\boldsymbol \xi) G_{\;\;\alpha}^{y}(\boldsymbol \xi)\right\}\ .
\end{equation}
Based on the same argument for the quark transverse AM, the twist-3 covariant relation for the transverse AM is equivalent to the the longitudinal one, which can be written as
\begin{equation}
\begin{split}
\label{jxgluont3}
J_{g}^{x(3)}=\frac{1}{2} \int \text{d}x\;x\; G_{g,3}(x)\ ,
\end{split}
\end{equation}
analogous to eq.~(\ref{jgzsum}). Again it has been renormalized with a factor of $1/2\gamma$ for simplicity. 

With the covariant gluon AM decomposition in section \ref{gluonamdecomp} and the definitions in eq.~(\ref{lgzdef})-eq.~(\ref{lgzpotdef}), similar matrix elements can be defined as
\begin{align}
\tilde L_g^{x(3)}&\equiv 2\int\text{d}^3\boldsymbol \xi\bra{P,S} \text{Tr}\left\{ G^{0 \alpha}\left(\xi^{z} \times D^{y}\right) A_{\alpha}\right\}\ket{P,S}\ ,\\
l_g^{x(3)}&\equiv2\int\text{d}^3\boldsymbol \xi\bra{P,S} \text{Tr}\left\{ G^{0 \alpha}\left(\xi^{z} \times \partial^{y}\right) A_{\alpha}\right\}\ket{P,S}\ ,\\
l_{g,\rm{pot}}^{x(3)}&\equiv-2ig\int\text{d}^3\boldsymbol \xi\bra{P,S} \text{Tr}\left\{  G^{0 \alpha}\left(\xi^{z} \times  A^{y}\right) A_{\alpha}\right\}\ket{P,S}\ ,
\end{align}
with $l_g^{x(3)}=\tilde L_g^{x(3)}+l_{g,\rm{pot}}^{x(3)}$ given by  eq.~(\ref{giegluon}). Then the twist-3 gluon AM can be expressed as
\begin{equation}
\label{j3gparton}
 J^{x(3)}_{g}=l_g^{x(3)}+l_{q,\rm{pot}}^{x(3)}+ \Delta G_T\ ,
\end{equation}
where the last term here defined as the $\Delta G_T\equiv 4\int\text{d}^3\boldsymbol \xi\bra{P,S} \text{Tr}\left\{G^{0 z}(\boldsymbol \xi) A^{y}(\boldsymbol \xi)\right\}\ket{P,S}$, which is
the same as $\Delta G$ because of the rotational 
symmetry. In the present case, it is naturally related to a twist-3 polarized gluon
distribution $\Delta G_T(x)$ ~\cite{Ji:1992eu,Kodaira:1998jn,Hatta2013} through $\Delta G_T 
= \int \Delta G_T(x) dx$,  where  
\begin{equation}
\begin{split}
\Delta G_T(x)=\frac{i}{ x P^+} \int \frac{\mathrm{d} \lambda}{ 2\pi} e^{i x \lambda}\bra{ P,S}2 \text{Tr}\left\{ G^{+\alpha}\left(-\frac{\lambda n}{2}\right)W_{-\frac{\lambda}{2},\frac{\lambda}{2}} \tilde{G}_{\;\;\alpha}^{\perp}\left(\frac{\lambda n}{2}\right) \right\}\ket{P,S}\ . 
\end{split}
\end{equation}
It is the transversely-polarized gluon density in a transversely-polarized 
nucleon, counterpart to the gluon helicity distribution in a longitudinally-polarized nucelon in eq.~(\ref{gluonhelicity}). 

Since $l_{q,\rm{pot}}^{x(3)}$ in eq.~(\ref{j3gparton}) cancels the term in the corresponding quark AM, the relation combining the quark decomposition eq.~(\ref{jqx3decomp}) immediately gives a
partonic sum rule in the transverse direction as in eq.~(\ref{jmsumrulex}) for the longitudinal spin. 

The gluon canonical OAM $l_g^{x(3)}$ can
be related to a parton density, 
\begin{equation}
l_g^{x(3)}(x)=\int \frac{\mathrm{d} \lambda\text{d}^2\boldsymbol \xi^\perp}{2 \pi}  \mathrm{e}^{i \lambda x}\left\langle P, S\right|2\text{Tr}\left\{G^{+\alpha}(0,\boldsymbol \xi^\perp) \xi^z \partial^y A_{\alpha}(\lambda n,\boldsymbol \xi^\perp)\right\}\left| P ,S\right\rangle\ ,
\end{equation}
as $l_g^{x(3)}=\int \text{d} x l_g^{x(3)}(x)$. The density $l_g^{x(3)}(x)$ has a simple parton interpretation as the canonical OAM carried
by a gluon with linear momentum $x P^+$. 
According to eq.~(\ref{GIEgluonOAM}), $l_{g}^{x(3)}(x)$ can be expressed with D- and F-type GPDs as 
\begin{equation}
    \begin{split}
        l_{g}^{x(3)}(x)= \int\text{d}y\bigg{[}&\text{P}\frac{1}{x+y}\left( G_{g,D,3}(x,y)+G'_{g,D,4}(x,y)\right)\\
        &+\text{P}\frac{1}{y^2-x^2}\left(G_{g,F,3}(x,y)+G'_{g,F,4}(x,y)\right)\bigg{]}\ .
    \end{split}
\end{equation}
Similarly, $\tilde L_g^{x(3)}$ and $l_{g,\rm{pot}}^{x(3)}$ can also be expressed with the D- and F-type GPDs as
\begin{align}
\tilde L_g^{x(3)}&= \int\text{d}x\text{d}y\text{P}\frac{1}{x+y}\left( G_{g,D,3}(x,y)+G'_{g,D,4}(x,y)\right)\ ,\\
l_{g,\rm{pot}}^{x(3)}&=  \int\text{d}x\text{d}y \text{P}\frac{1}{y^2-x^2}\left(G_{g,F,3}(x,y)+G'_{g,F,4}(x,y)\right)\ ,
\end{align}
which is consistent with relation $l_g^{x(3)}=\tilde L_g^{x(3)}+l_{g,\rm{pot}}^{x(3)}$.

Combining all the twist-3 transverse AM densities discussed so far, we arrive at the following sum rule 
\begin{equation}
\label{jmsumrulelf}
 \int \text{d} x\left(\frac{1}{2} g_T(x)+  \Delta G_{T}(x) + l_q^{x(3)}(x)+ l_g^{x(3)}(x )  \right)=\frac{\hbar}{2}\ ,
\end{equation}
with each term having a clear parton interpretation: $g_T(x)/\Delta G_{T}(x)$ gives the parton transverse spin densities, and $l_{q,g}^{x(3)}(x)$ are the parton transverse canonical OAM densities. These densities differ from those for the longitudinal AM because of the Lorentz boost, but their integrals are the same as the longitudinal ones by rotational symmetry. 

Finally, we mention that all terms in the above equation are renormalization scale dependent. Furthermore, when evolved from a particular scale, renormalization equations are complicated as they do not form a closed set: twist-three evolution generally involves light-cone correlations with two momentum fractions. 

\section{Transverse-space distributions of partons, AM and magnetic moment}
\label{sec5}
Although the twist-2 part of the transverse AM provides a simple partonic 
interpretation of the nucleon's transverse spin, it makes use of only the 
forward limit of the related GPDs. On the other hand, it was pointed out that 
GPDs with arbitrary $t$ but $\xi=0$ can be used to picture the 
transverse-space distributions of partons~\cite{Burkardt2003}.
This method helps to explore effects of the transverse polarization on 
the parton distributions in the transverse plane. However, 
as we learned in the previous section, not all effects 
related to the transverse polarization are intrinsic, it
is critically important to factor out the non-intrinsic contributions 
which are not related to the internal structure of the nucleon. 

In this section, we study the transverse-space distributions of partons inside the nucleon, 
particularly the effects from the transverse polarization. We find some
interesting results which are different from the standard formula in the 
literature~\cite{Burkardt2005,Miller:2010nz,Dahiya:2015jnn}. 
Our results help to understand interesting physical observables, such as OAM
and the magnetic moment~\cite{Miller:2007kt} in term of 3-dimensional images of the nucleon, 
or nuclear femtography.  
    
\subsection{Transverse-space distributions of partons in a transversely-polarized nucleon}

We begin by reviewing the relation between GPDs and transverse-space distributions proposed in ref. \cite{Burkardt2003,Burkardt2005}. In order to define
transverse coordinates $\boldsymbol b$ for partons, the nucleon state shall be taken as localized in the 
transverse plane through superimposing plane waves with all transverse momentum,
while taking its $z$-component going to $P^z=\infty$, 
\begin{equation}
    \left|P^z_\infty,\boldsymbol R^\perp=0\right> \equiv  \mathcal N \int \frac{d^2 \boldsymbol p^\perp}{(2\pi)^2} \left|P^z_\infty,\boldsymbol p^\perp\right> \ ,
\end{equation}
where $\left|P^z_\infty,\boldsymbol p^\perp\right>$ is a momentum eigen-state and $\mathcal N$ is a normalization constant. Using the above, we can define the quark distribution in the transverse space as, 
\begin{equation}
\label{unpqdistimp}
    \rho_{q}(x,\boldsymbol b)\equiv\int \frac{\mathrm{d} \lambda}{2 \pi}  \mathrm{e}^{i \lambda x}\bra{P^z_\infty,\boldsymbol R^\perp=0} \bar\psi\left(-\frac{\lambda n}{2},\boldsymbol b\right) \gamma^+ \psi\left(
\frac{\lambda n}{2},\boldsymbol b\right)\ket{P^z_\infty,\boldsymbol R^\perp=0}\ ,
\end{equation}
with gauge links between fields omitted. Then the Fourier transform of $\rho_{q}(x,\boldsymbol b)$, \begin{equation}
    \rho_{q}(x,\boldsymbol \Delta) \equiv \int \text{d}^2\boldsymbol b e^{i \boldsymbol{\Delta}\cdot \boldsymbol b}\rho_{q}(x,\boldsymbol b)  \ ,
\end{equation}
can be expressed as
\begin{equation}
\begin{split}
 \rho_{q}(x,\boldsymbol \Delta)= {\mathcal N}^2(2\pi)^2\delta^{(2)}(\boldsymbol 0) \int \frac{ \text{d}\lambda}{2\pi}e^{i \lambda  x}\bra{P'}\bar\psi\left(-\frac{\lambda n}{2}\right) \gamma^+ \psi\left(\frac{\lambda n}{2}\right)\ket{P}\ ,
\end{split}
\end{equation}
which is the twist-2 GPDs defined in eq.~(\ref{vectorGPD}) with $\Gamma=\gamma^+$ up to a normalization, 
except $\boldsymbol \Delta$ is in the transverse direction and hence $\xi=0$. 
With the parameterization in eq.~(\ref{gamma+exp}), the quark distribution can be expressed as
\begin{equation}
\label{qtransdistunp}
    \rho_{q}(x,\boldsymbol \Delta)={\mathcal N}^2(2\pi)^2\delta^{(2)}(\boldsymbol 0) \bar u(P')\bigg{(}H_q(x,-\boldsymbol \Delta^2) \gamma^+ 
    + E_q(x,-\boldsymbol \Delta^2) \frac{i\sigma^{+\alpha} \Delta_\alpha}{2M}\bigg{)}u(P)\ .
\end{equation}
In the infinite momentum frame where $P^z \to \infty$, it becomes simply
\begin{equation}
    \rho_{q}(x,\boldsymbol \Delta)=H_q(x,-\boldsymbol \Delta^2)\ ,
\end{equation}
with proper normalization, and the $E_q(x,-\boldsymbol \Delta^2)$ term does not contribute 
for an unpolarized state. Then the quark transverse space distribution 
and $H_q(x,-\boldsymbol \Delta^2)$ is related 
\begin{equation}
\label{densityfromgpd}
\begin{split}
    \rho_{q}(x,\boldsymbol b) = \int \frac{\text{d}^2\boldsymbol \Delta}{(2\pi)^2} e^{-i \boldsymbol{\Delta}\cdot \boldsymbol b}H_q(x,-\boldsymbol \Delta^2)    =\mathcal H_{q}(x,\boldsymbol b)\ ,
\end{split}
\end{equation}
where the Fourier transformation of GPD in transverse space $\mathcal H_{q}(x,\boldsymbol b)$ is defined. Based on its definition in eq.~(\ref{unpqdistimp}), $\rho_{q}(x,\boldsymbol b)$ is 
the transverse-plane distribution of quarks with momentum $x P^z$ in the localized 
nucleon state. 

If one sums over all partons, $\rho_{q}(\boldsymbol b) \equiv \int \text{d} x \rho(x,\boldsymbol b)$, expressed in a matrix element as
\begin{equation}
    \rho_{q}(\boldsymbol b)=\bra{P^z_\infty,\boldsymbol R^\perp=0} \bar\psi\left(\boldsymbol b\right) \gamma^+ \psi\left(
\boldsymbol b\right)\ket{P^z_\infty,\boldsymbol R^\perp=0}\ . 
\end{equation}
The quark density distribution in the transverse plane becomes  
\begin{equation}
    \rho_{q}(\boldsymbol b) =\int \frac{\text{d}^2\boldsymbol \Delta}{(2\pi)^2} e^{-i \boldsymbol{\Delta}\cdot \boldsymbol b}\int \text{d}x H_q(x,-\boldsymbol \Delta^2)= \int \frac{\text{d}^2\boldsymbol \Delta}{(2\pi)^2} e^{-i \boldsymbol{\Delta}\cdot \boldsymbol b}F_{1,q}(-\boldsymbol \Delta^2) \ ,
\end{equation}
which is the Fourier transformation of the Dirac form factor $F_{1,q}(-\boldsymbol \Delta^2)$ of the vector current~\cite{Soper:1976jc}. 
The above can be used to define charge distributions of the nucleon~\cite{Miller:2007uy,Miller2019}
without the relativistic effects affecting the standard interpretation of nucleon electromagnetic
form factors in the textbook~\cite{Sachs:1962zzc,Halzen:1984mc}.

As for transversely-polarized nucleons, the polarized state $\left|P^z,\boldsymbol R^\perp=0,S\right> $ needs a careful consideration. Since the states on the RHS use different $\boldsymbol p_\perp$, it is unclear how to have the same polarization vector $S$ in a transverse direction while satisfying the constraint $S\cdot P=0$ for all of them. However, in the infinite momentum limit where $P^z \to \infty$, all the states $\left|P^z,\boldsymbol p^\perp\right>$ with finite $\boldsymbol p^\perp$ can be in a definite helicity state, $h=\pm$. Then the transversely-polarized nucleon can be defined as \cite{Burkardt2003,Miller:2007kt},
\begin{equation}
    \left|P^z,X\right> \equiv \frac{1}{\sqrt{2}}\left(\left|P^{z}, \mathbf{R}^\perp=0,+\right\rangle+\left|P^{z}, \mathbf{R}^\perp=0,-\right\rangle\right)\ ,
\end{equation}
where $\left|P^{z}, \mathbf{R}^\perp=0,+(-)\right\rangle$ are states localized in the transverse plane with  definite helicity. The transverse-space quark density distribution in polarized nucleon states can be defined as
\begin{equation}
\label{pqdistimp}
    \rho_{q}^{X}(x,\boldsymbol b)\equiv \int \frac{\mathrm{d} \lambda}{2 \pi}  \mathrm{e}^{i \lambda x}\bra{P^z,X} \bar\psi\left(-\frac{\lambda n}{2},\boldsymbol b\right) \gamma^+ \psi\left(
\frac{\lambda n}{2},\boldsymbol b\right)\ket{P^z,X}\ .
\end{equation}
Its Fourier transformation is then
\begin{equation}
\begin{split}
 \rho_{q}^{X}(x,\boldsymbol \Delta)=\frac{1}{2}{\mathcal N}^2(2\pi)^2\delta^{(2)}(\boldsymbol 0)\sum_{h',h}\int \frac{ \text{d}\lambda}{2\pi}e^{i \lambda  x}\bra{P',h'}\bar\psi\left(-\frac{\lambda n}{2}\right) \gamma^+ \psi\left(\frac{\lambda n}{2}\right)\ket{P,h}\ ,
\end{split}
\end{equation}
which is related to the twist-2 GPDs,
\begin{equation}
    \rho_{q}^{X}(x,\boldsymbol \Delta)\propto \frac{1}{2}\sum_{h',h}\bar u(P',h')\bigg{(}H_q(x,-\boldsymbol \Delta^2) \gamma^+ 
    + E_q(x,-\boldsymbol \Delta^2) \frac{i\sigma^{+\alpha} \Delta_\alpha}{2M}\bigg{)}u(P,h)\ ,
\end{equation}
where summations over helicity are performed for the initial and final states, separately.
Using the spinors in the infinite momentum limit $P^z\to \infty$, one then finds 
\begin{equation}
    \rho_{q}^{X}(x,\boldsymbol \Delta)=H_q(x,-\boldsymbol \Delta^2)+\frac{i\Delta_y}{2M}E_q(x,-\boldsymbol \Delta^2)\ .
\end{equation}
After inverse Fourier-transformed to the transverse space, 
this gives the quark distribution in transversely polarized nucleons  
\begin{equation}
\begin{split}
\label{rhoXb}
    \rho_{q}^{X}(x,\boldsymbol b)=&\int \frac{\text{d}^2\boldsymbol \Delta}{(2\pi)^2} e^{-i \boldsymbol{\Delta}\cdot \boldsymbol b} \left(H_q(x,-\boldsymbol \Delta^2)+\frac{i\Delta_y}{2M}E_q(x,-\boldsymbol \Delta^2)\right)\ ,\\
    =& \mathcal H_q(x,\boldsymbol b) -\frac{1}{2M}\frac{\partial}{\partial b^y} \mathcal E_q(x,\boldsymbol b)\ ,
\end{split}
\end{equation}
with the Fourier-transformed GPDs $\mathcal H_q(x,\boldsymbol b)/ \mathcal E_q(x,\boldsymbol b)$ in transverse space, 
as first derived in ref. \cite{Burkardt2003}.  

As discussed in the last section, however, the result above also receives contributions from the 
center-of-mass motion which needs to be removed to get the intrinsic quark distributions. To show that, 
the following spinor expansions in the infinite momentum limit will be useful
\begin{align}
    \frac{1}{2}\sum_{h',h}\bar u(P',h')u(P,h) &=2M+\epsilon^{0\rho\alpha\beta} \frac{i\Delta_\rho}{\bar P^0+M}S_\alpha \bar P_\beta\ ,\\
    \frac{1}{2}\sum_{h',h}\bar u(P',h')\sigma^{\mu\nu} u(P,h) &=- 2\epsilon^{\mu\nu\alpha\beta}S_\alpha \bar P_{\beta}\ ,
\end{align}
with the $S_\alpha$ polarization vector in the $x$-direction. The expansions above coincide with those 
in eq.~(\ref{spinorscalar}) and eq.~(\ref{sigmaspinor}), because all plane-wave states 
in the localized nucleon $\left|P^z,X\right>$ have the same $P^z\to \infty$ and the momentum
difference $\boldsymbol \Delta$ is in the transverse direction. 
In the infinite momentum limit, with the spinor expansions above, $\rho_{q}^{X}(x,\boldsymbol \Delta)$ can be expressed as
\begin{equation}
\begin{split}
\label{currentdist}
    \rho_{q}^{X}(x,\boldsymbol \Delta)=&\frac{\bar P^\mu}{\bar P^0}H_q(x,-\boldsymbol \Delta^2)\left(1+\epsilon^{0\rho\alpha\beta} \frac{i\Delta_\rho}{2M(\bar P^0+M)}S_\alpha \bar P_\beta\right)\\
    &-\left(H_q(x,-\boldsymbol \Delta^2)+E_q(x,-\boldsymbol \Delta^2)\right)\frac{i\epsilon^{\mu\nu\alpha\beta}\Delta_\nu}{2M \bar P^0} S_\alpha \bar P_\beta\ .
\end{split}
\end{equation}

However, the second term in the first line of eq.~(\ref{currentdist}), the $\epsilon^{0\rho\alpha\beta}$ term, corresponds to a contribution to the center-of-mass motion induced by the transverse polarization, as discussed in the last section. It nominally cancels the $H_q(x,-\boldsymbol \Delta^2)$ term in the second line, but has a completely different
Lorentz transformation property. 
According to the definition of the Pauli-Lubanski spin vector, such term 
shall not contribute to the intrinsic 
spin property of the nucleon because it is
proportional to the center-of-mass motion. 
It shall be removed so the transverse AM restores the proper $\gamma$ factor as indicated by its Lorentz transformation~\cite{Ji2020}. Therefore, it should be also be removed from 
the intrinsic charge density. We then define 
the intrinsic quark density distribution in 
transverse-space through,
\begin{equation}
    \rho_{q,\rm{In}}^{X}(x,\boldsymbol \Delta)=H_q(x,-\boldsymbol \Delta^2)+\frac{i\Delta_y}{2M}\left(H_q(x,-\boldsymbol \Delta^2)+E_q(x,-\boldsymbol \Delta^2)\right)\ .
\end{equation}
And its Fourier transformation becomes 
\begin{equation}
\begin{split}
\label{rhoinXb}
    \rho_{q,\rm{In}}^{X}(x,\boldsymbol b)=&\int \frac{\text{d}^2\boldsymbol \Delta}{(2\pi)^2} e^{-i \boldsymbol{\Delta}\cdot \boldsymbol b} \left[H_q(x,-\boldsymbol \Delta^2)+\frac{i\Delta_y}{2M}\left(H_q(x,-\boldsymbol \Delta^2)+E_q(x,-\boldsymbol \Delta^2)\right)\right]\ , \\
    =&  \mathcal H_q(x,\boldsymbol b)-\frac{1}{2M}\frac{\partial}{\partial b^y} \left( \mathcal H_q(x,\boldsymbol b)+ \mathcal E_q(x,\boldsymbol b) \right) \ ,
\end{split}
\end{equation}
which differs from eq.~(\ref{rhoXb}) by the $\mathcal H_q(x,\boldsymbol b)$ term. 

\subsection{Distribution of transverse AM in transverse space}

With the intrinsic parton distribution $\rho_{q,\rm{In}}^{X}(x,\boldsymbol b)$, the transverse AM density in the transverse plane can be derived straightforwardly. The twist-2 quark transverse AM density in the transverse plane can be defined as
\begin{equation}
    J^{x(2)}_{q}(x,\boldsymbol b)\equiv \int \frac{\mathrm{d} \lambda}{2 \pi}  \mathrm{e}^{i \lambda x}\bra{P^z,X} \bar\psi\left(-\frac{\lambda n}{2},\boldsymbol b\right) \gamma^+  b^y i\overleftrightarrow{\partial}^+ \psi\left(
\frac{\lambda n}{2},\boldsymbol b\right)\ket{P^z,X}\ ,
\end{equation}
whose inverse Fourier transformation is
\begin{equation}
\label{jx2momgpd}
\begin{split}
    J^{x(2)}_{q}(x,\boldsymbol \Delta)&= ( b^y \times  x P^+)\rho_{q,\rm{In}}^{X}(x,\boldsymbol \Delta)\ ,\\
    &=\frac{\gamma}{2}x\left(H_q(x,-\boldsymbol \Delta^2)+E_q(x,-\boldsymbol \Delta^2)\right)\ ,
\end{split}
\end{equation}
where $b^y$ becomes the derivative operator $-i\frac{\partial}{\partial \Delta_y}$ in momentum representation and $( b^y \times  x P^+)$ is the twist-2 transverse AM operator. In the forward limit, it is equivalent to the transverse AM derived in eq.~(\ref{jx2qlf}). The quark transverse AM distribution in the transverse space $J^{x(2)}_{q}(x,\boldsymbol b)$ is now
\begin{equation}
\label{jx2gpd}
    J^{x(2)}_{q}(x,\boldsymbol b)=\frac{\gamma}{2}x \left(\mathcal H_q(x,\boldsymbol b)+\mathcal E_q(x,\boldsymbol b)\right)\ ,
\end{equation}
through Fourier transformation.

The total quark transverse AM distribution in the transverse space defined as
\begin{equation}
    J^{x(2)}_{q}(\boldsymbol b)\equiv \bra{P^z,X} \bar\psi\left(\boldsymbol b\right) \gamma^+  b^y i\overleftrightarrow{\partial}^+ \psi\left(
\boldsymbol b\right)\ket{P^z,X}\ , 
\end{equation}
can then be shown to be simply the sum over all quarks of different $x$ as $J^{x(2)}_{q}(\boldsymbol b) = \int \text{d} x J^{x(2)}_{q}(x,\boldsymbol b) $. With eq.~(\ref{jx2momgpd}), $J^{x(2)}_{q}(\boldsymbol b)$ can be expressed as
\begin{equation}
    J^{x(2)}_{q}(\boldsymbol b)=\frac{\gamma}{2}\int \frac{\text{d}^2\boldsymbol \Delta}{(2\pi)^2} e^{-i \boldsymbol{\Delta}\cdot \boldsymbol b} \left(A_q(-\boldsymbol \Delta^2)+B_q(-\boldsymbol \Delta^2)\right)\ ,
\end{equation}
which relates the gravitational form factors $A_q(-\boldsymbol \Delta^2)$ and $B_q(-\boldsymbol \Delta^2)$ to the AM distribution in the transverse plane of the nucleon. The gluon AM distribution can be defined similarly and the total AM density is $J^{x(2)}(\boldsymbol b)\equiv \sum_i J^{x(2)}_{i}(\boldsymbol b)$ with $i$ summing over quarks and gluons.

\subsection{Distribution of magnetic moment in transverse space}

Distributions of other quantities in transversely-polarized nucleons can be derived similarly, such as the magnetic moment whose property resembles that of AM. The magnetic moment can be written as
\begin{equation}
    \boldsymbol m =\frac{1}{2}\int \text{d}^3 \boldsymbol r \;\bra{P}\boldsymbol r \times \boldsymbol {\mathcal J}_{EM}\ket{P}\ ,
\end{equation}
with ${\mathcal J}_{EM}$ the electromagnetic current of the system which simply sums over all the quark currents of different flavors $i$ and charges $e_i$ in nucleon as
\begin{equation}
\begin{split}
\label{emcurrent}
{\mathcal J}_{EM}^\mu=\sum_i e_i {\mathcal J}_{i}^\mu=\sum_i e_i \bar\psi_i \gamma^\mu \psi_i\ .
\end{split}
\end{equation}
The magnetic moment of nucleons can be related to their spin $\boldsymbol s$ as 
\begin{equation}
    \boldsymbol m =\mu \boldsymbol s \ ,
\end{equation}
so the scalar $\mu$ also gives the magnetic moment of nucleons. For a transversely polarized nucleon in momentum eigen-states, the magnetic moment can be written as
\begin{equation}
    \boldsymbol m^x =\frac{1}{2}\int \text{d}^3 \boldsymbol r \;\bra{P,S}\left(\boldsymbol r^y \boldsymbol {\mathcal J}_{EM}^z-\boldsymbol r^z \boldsymbol {\mathcal J}_{EM}^y\right)\ket{P,S}\ .
\end{equation}
The same as AM, in the infinite momentum frame, it can be split into the twist-2 $\boldsymbol r^y\boldsymbol {\mathcal J}_{EM}^z$ term and twist-3 $-\boldsymbol r^z\boldsymbol {\mathcal J}_{EM}^y$ term. The twist-2 magnetic moment which dominates in the infinite momentum limit is defined as
\begin{align}
    \mu^{(2)}\equiv\bra{P^z,S}\int \text{d}^3 \boldsymbol r \;\boldsymbol r^y \boldsymbol {\mathcal J}_{EM}^z\ket{P^z,S}\ ,
\end{align}
Just like the AM which can be expressed with the gravitational form factors, the magnetic moment can be related to the electromagnetic form factor which is defined as,
\begin{equation}
    \bra{P'} {\mathcal J}^{\mu}_{EM}\ket{P}= e_N \bar u(P')\left(F_{1}(\Delta^2) \gamma^\mu +F_{2}(\Delta^2)\frac{i\sigma^{\mu\nu}\Delta_\nu}{2M}\right) u(P)\ ,
\end{equation}
where $e_N$ is the charge of the nucleon and $F_{1}(\Delta^2)$ and $F_{2}(\Delta^2)$ are the Dirac and Pauli form factors of nucleon. Then the magnetic moment $\mu$ can be expressed as
\begin{equation}
\label{magnetff}
   \mu=\frac{e_N \hbar}{2M} (F_1(0)+F_2(0))\ ,
\end{equation}
which can also be expressed with the nuclear magneton $\mu_N\equiv e_N\hbar/(2M)$ and the magnetic form factor $G_{M}(\Delta^2) \equiv F_1(\Delta^2)+F_2(\Delta^2)$ as $\mu=\mu_N G_{M}(0)$.

The magnetic moment distribution in the transverse space follows from   
\begin{equation}
    \mu^{(2)}(x,\boldsymbol b)=  b^y \int \frac{\mathrm{d} \lambda}{2 \pi}  \mathrm{e}^{i \lambda x}\bra{P^z_\infty,X} \sum_i e_i \bar\psi_i\left(-\frac{\lambda n}{2},\boldsymbol b\right) \gamma^+  \psi_i\left(
\frac{\lambda n}{2},\boldsymbol b\right)\ket{P^z_\infty,X}\ .
\end{equation}
Again, with the intrinsic parton distribution in transverse space $\rho_{q_i,\rm{In}}^{X}(x,\boldsymbol b)$, one has the simple relation
\begin{equation}
    \mu^{(2)}(x,\boldsymbol \Delta)=  b^y\sum_i e_i \rho_{q_i,\rm{In}}^{X}(x,\boldsymbol \Delta)\ ,
\end{equation}
in momentum representation with again $b^y$ the derivative operator $-i\frac{\partial}{\partial \Delta_y}$.
Using eq.~(\ref{rhoinXb}), it can be written with GPDs in transverse space as
\begin{equation}
    \mu^{(2)}(x,\boldsymbol b)= \sum_i\frac{e_i\hbar}{2M}\left(\mathcal H_{q_i}(x,\boldsymbol b)+ \mathcal E_{q_i}(x,\boldsymbol b)\right)\ ,
\end{equation}
where $-i\frac{\partial}{\partial \Delta_y}$ takes the term proportional to $\Delta_y$ only. The relation connects the magnetic moment density distribution to GPDs in transverse space, analogous to eq.~(\ref{jx2gpd}) for AM.

The total magnetic moment distribution in the transverse space defined as
\begin{equation}
    \mu^{(2)}(\boldsymbol b)=  b^y \bra{P^z,X} \sum_i e_i \bar\psi_i\left(\boldsymbol b\right) \gamma^+  \psi_i\left(
\boldsymbol b\right)\ket{P^z,X}\ ,
\end{equation}
is then a sum over all partons: $\mu^{(2)}(\boldsymbol b)= \int \text{d}x \mu^{(2)}(x, \boldsymbol b)$. Following relation between $\mu^{(2)}(\boldsymbol b)$ and electromagnetic form factors can be derived as
\begin{equation}
\begin{split}
    \mu^{(2)}(\boldsymbol b)=\sum_i \frac{e_i\hbar}{2M}\int \frac{\text{d}^2\boldsymbol \Delta}{(2\pi)^2} e^{-i \boldsymbol{\Delta}\cdot \boldsymbol b} \left(F_{1,q_i}(-\boldsymbol \Delta^2)+F_{2,q_i}(-\boldsymbol \Delta^2)\right)\ ,
\end{split}
\end{equation}
which indicates that the magnetic form factors $G_{M}(-\boldsymbol\Delta^2)$ describe the magnetic moment distribution, analogous to the relation between Dirac form factor and charge distribution. This result agrees with the form factor result in eq.~(\ref{magnetff}). It confirms the previous result~\cite{Miller:2007kt} for the anomalous magnetic momentum distribution in the transverse space.

\section{Conclusion}
\label{sec6}

In this work we reviewed the known spin structure theory of the nucleon, while breaking some important and new ground on the subject.  Deriving spin sum rules for the nucleon comes with some important factors that must be considered, including: frame choice, decomposition choice of AM contributions, the polarization of the nucleon, twist-counting, and isolating intrinsic contributions.  By carefully heeding these factors, meaningful quantitative statements about the spin structure of the nucleon are derived.    


In this paper we have defined an extensive list of twist-2 and twist-3 GPDs related to QCD energy-momentum tensor, some of which involve a more general Lorentz structure than has been considered in the past. This was done with the goal of studying the intrinsic AM of the nucleon in great detail.  We have also stated the mathematical properties of these distributions and laid out many of their interrelations, with some connections to previously defined GPDs included.  

We have recovered both the covariant and infinite-momentum-frame AM sum rules of the nucleon. 
The covariant sum rule has simple partonic interpretation at leading twist in transverse polarization only. While for the longitudinal polarization, a decomposition of AM onto spin and orbital contribution and the infinite-momentum-frame sum rule is needed to have a simple partonic interpretation as first done by Jaffe and Manohar \cite{JaffeMan90}. We have successfully applied this concept to the twist-3 contribution of the transverse AM in this paper. 

After analyzing the twist-3 contribution to transverse AM, we find a new partonic sum rule for a transversely-polarized nucleon in eq.~(\ref{jmsumrulelf}), which involves the known transverse spin distributions of the quarks and gluons as well as some novel twist-3 GPDs associated with the canonical OAM of the partons. This sum rule is in some sense a ``rotated" version of the Jaffe-Manohar sum rule for longitudinal polarization. While each term in the transverse sum rule equals to the corresponding one in the longitudinal case after summing over all partons, they are distinguished by their different partonic densities.


At last we studied the transverse plane distributions of partons and other quantities using GPDs with non-zero transverse momentum transfer. We derived expressions for the intrinsic quark density distribution, the transverse AM distribution of quarks, and the magnetic moment distribution in transverse space, all in terms of twist-2 GPDs.  
The results derived differ from the ones in the literature, highlighting the distinction between intrinsic and extrinsic contributions.

 Looking forward, a reasonable goal is to experimentally complete the twist-2 sum rule with new data which can constrain the involved twist-2 GPDs.  Meanwhile, a computation of the GPDs which completes the twist-3 sum rules should also be examined.  A third ambitious goal would be to experimentally constrain the twist-3 AM GPDs.  With that in mind, we are now tasked with relating our defined leading and sub-leading GPDs found in these sum rules to the experimental cross sections planned in upcoming experiments such as those for the DVCS processes planned at the EIC.  Some impressive work has been done on this topic \cite{Kriesten:2019jep,Belitsky:2001ns,Kiptily2002}, however some discrepancies exist between the authors and we believe that a further examination is warranted, and one which is consistent with our formalism used in this paper.

\section*{Acknowledgments}
We thank M. Burkardt, Y. Hatta, S. Liuti, F. Yuan, and Y. Zhao for discussions related to the subject of this paper. This research is supported by the U.S. Department of Energy, Office of Science, Office of Nuclear Physics, under contract number DE-SC0020682, 
and the Center for Nuclear Femtography, Southeastern Universities Research Association, Washington D.C.
\newpage
\appendix

\section{Defining AM}
\label{defam}
We begin our discussion of AM by looking at the EMT.  The EMT is a fundamental object which plays a role in both classical and quantum field theory and can be extracted from the system Lagrangian density itself 
\begin{equation}\label{EMTgeneral}
    T^{\mu\nu}(\xi)=\frac{\partial\mathcal{L}}{\partial(\partial_\mu \phi)}\partial^\nu \phi(\xi) - g^{\mu\nu}\mathcal{L}(\xi)\ ,
\end{equation}
and is a conserved quantity $\partial_\mu T^{\mu\nu}=0$ as well as the generator of the covariant momentum $P^\mu$ of a particle.  It has two standard definitions however, namely the canonical eq.~(\ref{EMTgeneral}) and the Belinfante versions.  The Belinfante version is constructed to be both conserved and symmetric, and leads to the same conserved charge as the canonical definition
\begin{equation}\label{Belifante}
    T_{\text{Bel}}^{\mu\nu}(\xi)=T^{\mu\nu}(\xi) + \partial_\lambda G^{[\lambda\mu]\nu}(\xi)\ ,
\end{equation}
where the $G^{[\lambda\mu]\nu}$ term is sometimes referred to as a \textit{superpotential}. Its specific form is not important, as in observables the term itself gets discarded as a surface term, and some care has been taken in the past to ensure this is valid in a wave packet analysis \cite{SHORE2000,Bakker2004}. We use eq.~(\ref{Belifante}) in both section 3 and 4, as it will be used to exploit a symmetry of contributions to the proton's OAM.
From the EMT we define the AM density operator as follows
\begin{equation}\label{AMdensity}
    M^{\mu\nu\rho}(\xi)=\xi^\nu T_{\text{Bel}}^{\mu\rho}(\xi) - \xi^\rho T_{\text{Bel}}^{\mu\nu}(\xi)\ ,
\end{equation}
which is again a conserved quantity $\partial_\mu M^{\mu\nu\lambda}=0$ and its serves as the generator of the relativistic AM matrix
\begin{equation}\label{AMmatrix}
    J^{\mu\nu}=\int d^3\boldsymbol{\xi}M^{0\mu\nu}(\boldsymbol \xi)\ .
\end{equation}
These matrix elements are a key ingredient in the so-called Pauli-Lubanski vector, which is a covariant 4-vector which can be used to represent the spin states of a relativistic particle.  It is formally defined by
\begin{equation}\label{PVvector}
    W^\mu = -\frac{1}{2}\epsilon^{\mu\alpha\lambda\sigma}\frac{J_{\alpha\lambda}P_\sigma}{M}\ ,
\end{equation}
where $\epsilon^{0123}=1$ and $P_\sigma$ is the 4-momentum of the particle.  From eq.~(\ref{PVvector}) follows a frame-independent spin scalar operator $W^\mu W_\mu$ with total spin eigenvalues
\begin{equation}
    W^\mu W_\mu \xrightarrow{\text{eigenvalues}} -s(s+1)\hbar^2\ .
\end{equation}
Returning to eq.~(\ref{AMmatrix}), both the boost operator and traditional AM operators in 3-space are extracted from it via
\begin{eqnarray}
K^i&=&J^{0i}\ ,\\
J^i&=&\frac{1}{2}\epsilon^{0ikl}J_{kl}\ .\label{AMop}
\end{eqnarray}
In high energy scattering, it is conventional to express our particle states with their simultaneous 4-momentum and spin polarization generically by $|P,S\rangle$ and with the normalization of $\langle P,S|P,S\rangle =2E_P(2\pi)^3\delta^3(0)$.  In a basis-independent fashion, a particle's momentum and polarization 4-vectors can be represented as follows
\begin{eqnarray}
    P^\mu &=& M\gamma (1,\vec{\beta})\ ,\\
    S^\mu &=& (\gamma \vec{s}\cdot\vec{\beta}\ , \vec{s}+(\gamma-1)(\vec{s}\cdot\hat{\beta})\hat{\beta})\ ,\label{spinpol}
\end{eqnarray}
where $\gamma=(1-|\vec{\beta}|^2)^{-\frac{1}{2}}$, $\vec{\beta}=\vec{v}/c$ and ensures that $P\cdot S=0$.  In the same spirit, the Pauli-Lubanski vector in eq.~(\ref{PVvector}) can alternatively be expressed as
\begin{equation}\label{Wu}
    W^\mu = \gamma(\vec{J}\cdot\vec{\beta},\vec{J}+\vec{K}\times\vec{\beta})\ .
\end{equation}
From eq.~(\ref{spinpol}) \& eq.~(\ref{Wu}), one can then define a generalized spin projection operator $S_p$ with their inner product
\begin{eqnarray}
    S_p&=&-W^\mu S_\mu\\
    &=&\gamma \vec{J}\cdot\vec{s}-(\gamma-1)\vec{s}\cdot{\beta}\vec{J}\cdot\vec{\beta}-\gamma \vec{s}\cdot(\vec{K}\times\vec{\beta})\label{eq13}\ .
\end{eqnarray}
It then follows that $|P,S\rangle$ is an eigenstate of $S_p$
\begin{equation}\label{Speigen}
    S_p|P,S\rangle = \frac{\hbar}{2}|P,S\rangle\ .
\end{equation}
Although eq.~(\ref{Speigen}) looks like an ideal equation to study the spin structure of the proton, because the $S_p$ operator includes both AM and boost operators eq.~(\ref{eq13}), disentangling them is a precarious task.  This was demonstrated in ref. \cite{Ji2020}, and for this reason we will be looking directly at the matrix elements of the AM operator in eq.~(\ref{AMop}).

Without loss of generality, we take the proton to be traveling in the $\hat{z}$ direction: $P^\mu=(P^0,0,0,P^z)$.  Transverse then refers to any direction in the $xy$ plane, while longitudinal is in the $z$ direction.
  Using the definitions above, we have the following forward expectation value for $J^x(\xi^0=0)=J^{yz}(\xi^0=0)$:
\begin{equation}\label{Jxearly}
 \langle P,S|J^x(\xi^0=0)|P,S\rangle = \int d^3\boldsymbol{\xi}\langle P,S|\xi^y T_{\text{Bel}}^{0z}(0,\boldsymbol{\xi})-\xi^z T_{\text{Bel}}^{0y}(0,\boldsymbol{\xi})|P,S\rangle.
\end{equation}
Keeping in mind that $P\cdot S=0$ for a physical nucleon state,  eq.~(\ref{Jxearly}) can be re-written in terms of an off-forward matrix element as
\begin{eqnarray}
 \langle P,S|J^x(\xi^0=0)|P,S\rangle &=& \lim_{\boldsymbol{\Delta}\to 0}\int d^3\boldsymbol{\xi}\langle P+\Delta, S'| \xi^y T_{\text{Bel}}^{0z}(0,\boldsymbol{\xi})-\xi^z T_{\text{Bel}}^{0y}(0,\boldsymbol{\xi})|P,S \rangle\nonumber\ , \\
 &=& \lim_{\boldsymbol{\Delta}\to 0}\int d^3\boldsymbol{\xi}e^{-i\boldsymbol{\Delta}\cdot\boldsymbol{\xi}}\langle P+\Delta, S'| \xi^y T_{\text{Bel}}^{0z}(0)-\xi^z T_{\text{Bel}}^{0y}(0)|P,S \rangle\ .\label{Jxinter}\nonumber\\
\end{eqnarray}
Due to the mass shell relations, in the limit as $\boldsymbol{\Delta}\rightarrow 0$ we also have that $\Delta^0\rightarrow 0$.  For each term in eq.~(\ref{Jxinter}) we can make the replacement $\xi^j e^{-i\boldsymbol{\Delta}\cdot\boldsymbol{\xi}}\rightarrow i\frac{\partial}{\partial\Delta^j}e^{-i\boldsymbol{\Delta}\cdot\boldsymbol{\xi}}$.  Thus we have the relation
\begin{equation}
 \langle P,S|J^x(\xi^0=0)|P,S\rangle = \lim_{\Delta\to 0}\int d^3\boldsymbol{\xi}\; \bigg{(}i \frac{\partial}{\partial\Delta^y} e^{-i\boldsymbol{\Delta}\cdot\boldsymbol{\xi}}\bigg{)} \langle P+\Delta, S'|T_{\text{Bel}}^{0z}(0)|P,S\rangle - (z \leftrightarrow y)\ .
 \label{JxIBP}
\end{equation}
From here we can use integration by parts to replace eq.~(\ref{JxIBP}) with a term which has the derivative acting on the whole expression and a term which has the derivative acting only on the matrix element.  The latter of these represents an internal AM contribution to the nucleon.  The former was shown to be the AM of the nucleon's wave packet about the origin \cite{Bakker2004} and we will consequently discard it.  The remaining $\xi$ dependence then resides solely in the complex exponential and we can use $\int d^3 \boldsymbol{\xi}e^{-i\boldsymbol{\xi}\cdot{\boldsymbol\Delta}}|_{\boldsymbol{\Delta}\rightarrow 0} = (2\pi)^3\delta^{(3)}(\boldsymbol 0)$.  We are then left with
\begin{equation}\label{Jxfinal}
   \langle P,S|J^x(\xi^0=0)|P,S\rangle = i(2\pi)^3\delta^{(3)}(0) \lim_{\Delta\to 0}\bigg{[} \frac{\partial}{\partial\Delta^y} \langle P+\Delta, S'|T_{\text{Bel}}^{0z}(0)|P,S\rangle - (z \leftrightarrow y) \bigg{]}\ .  
\end{equation}
The divergent prefactor will be cancelled in a properly normalized expectation value.  Eq.~(\ref{Jxfinal}) is an important result, as it relates our sought transverse AM matrix element to an off-forward matrix element of the EMT - which is well-defined in QCD.  The corresponding expression for the  $J^y$ and $J^z$ cases involve a trivial permutation between the $x$, $y$ and $z$ indices.

\section{Parameterization for GPDs}
\label{paramapdx}
Here it will be shown how our parameterization will form a complete basis in terms of Dirac structure for different types of GPDs. See similar discussion in ref. \cite{Meisner2009} and the Appendix of ref. \cite{Diehl2001}. The general strategy is to first eliminate as many Dirac structures as possible and then show that the remaining structures are independent and complete. To do so, the following Gordon identities 
\begin{align}
\label{gordon1}
2 M \bar{u}\left(P',S'\right) \gamma^{\alpha} u(P,S) &=\bar{u}\left(P',S'\right)\left[2\bar P^{\alpha}+i \sigma^{\alpha \beta}\Delta_{\beta}\right] u(P,S)\ , \\
\label{gordon2}
2 M \bar{u}\left(P',S'\right) \gamma^{\alpha} \gamma_{5} u(P,S), &=\bar{u}\left(P',S'\right)\left[\gamma_{5}\Delta^{\alpha}+2i \sigma^{\alpha \beta} \gamma_{5}\bar P_{\beta}\right] u(P,S)\ ,
\end{align}
and
\begin{align}
\label{gordon3}
\bar{u}\left(P',S'\right)\left[ \Delta^{\alpha}+2 i \sigma^{\alpha \beta}\bar P_{\beta}\right] u(P,S)=&0\ ,\\
\label{gordon4}
\bar{u}\left(P',S'\right)\left[2\bar P^{\alpha} \gamma_5+ i \sigma^{\alpha \beta}\gamma_5 \Delta_{\beta}\right] u(P,S)=&0\ ,
\end{align}
can be helpful. Eq.~(\ref{gordon1}) and eq.~(\ref{gordon2}) indicate that all the Dirac vectors $\gamma^\alpha$ and Dirac pseudo-vectors $\gamma^{\alpha} \gamma_5$ can be replaced with a Dirac scalar and a Dirac tensor $\sigma^{\alpha\beta}$ term. Moreover, another identity for the Dirac pseudo-tensor
\begin{equation}
    \sigma^{\alpha\beta}\gamma_5=-\frac{i}{2}\epsilon^{\alpha\beta\mu\nu}\sigma_{\mu\nu}\ ,
\end{equation}
and the projection of eq.~(\ref{gordon4}) to $n_\alpha$ which gives 
\begin{equation}
\begin{split}
    \bar{u}\left(P',S'\right)\left[2 \gamma_5+ i \sigma^{\alpha \beta}\gamma_5 n_\alpha \Delta_{\beta}\right] u(P,S)=&0\ ,\\
    \to\bar{u}\left(P',S'\right)\left[\gamma_5+\frac{1}{4}\epsilon^{\alpha\beta\mu\nu} n_\alpha \Delta_{\beta}\sigma_{\mu\nu}\right] u(P,S)=&0\ ,
\end{split}
\end{equation}
will be helpful as well.  They indicate that all Dirac pseudo-scalars and pseudo-tensors can be expressed with Dirac scalars and Dirac tensors, which are the only two kinds of Dirac structures remaining. In addition, eq.~(\ref{gordon3}) implies Dirac tensors in the form of $i \sigma^{\alpha \beta}\bar P_{\beta}$ will reduce to Dirac scalars. Thus in general, one has
\begin{equation}
   \text{all Dirac structures} \sim \text{Dirac scalar}+\text{Dirac tensor except with } \sigma^{\alpha \beta}\bar P_{\beta}\ .
\end{equation}
With this argument in mind, parameterization for different Dirac structures can be done.

\subsection{Parameterization of vector and tensor combination}

The starting point are those vectors without Dirac matrices, for which one has
\begin{equation}
    V_S^{\mu}=\{\bar P^\mu,\Delta^\mu,n^\mu\}\ ,
\end{equation}
whenever a GPD is concerned. The $S$ (scalar) subscript representing it is Dirac scalar contribution and then more complicated Dirac structure can be constructed by combining $V_S^{\mu}$ with Dirac matrices, while all the scalars will be absorbed into the coefficient in principle.

The decomposition must satisfy all the symmetries the origin structure has, including both Lorentz symmetry and the discrete symmetry such as parity and time reversal symmetry. An important constraint from the parity is that since all the Dirac pseudo-scalars, pseudo-vectors and pseudo-tensors are expressed with the other Dirac structures, there should be no $\epsilon^{\mu\nu\rho\sigma}$ term whose existence breaks the parity symmetry without those Dirac pseudo structures.

First, consider all the possible combinations that behave like Lorentz vectors: \newline
$\bar{u}\left(P',S'\right) \Gamma^\mu u(P,S)$\ .  The contribution from Dirac scalars is simple, since the only Dirac scalar is the identity matrix $I$. Thus the possible terms are just $V_S^{\mu}$, and one has:
\begin{equation}
   \Gamma^\mu\text{ with Dirac scalar} \sim V^\mu_S= \{P^\mu,\Delta^\mu,n^\mu\}\ .
\end{equation}
As for the tensor contribution, there are in principle two possible combinations:
\begin{equation}
   \Gamma^\mu \text{ with Dirac tensor} \sim \{\sigma^{\mu\nu} V_{S\nu}, V_S^\mu\sigma^{\alpha\beta} V_{S\alpha}V_{S\beta} \}\ .
\end{equation}
For future convenience, it is helpful to define
\begin{equation}
   V_T^\mu=\{\sigma^{\mu\nu} \Delta_{\nu},\sigma^{\mu\nu} n_{\nu}\}\ ,
\end{equation}
which stands for all possible vector combination with Dirac tensors, and
\begin{equation}
   S_T=\sigma^{\alpha\beta} n_{\alpha}\Delta_{\beta}\ ,
\end{equation}
which represents all possible scalar combinations with Dirac tensors. All the $\sigma^{\alpha\beta} \bar P_{\beta}$ are neglected which reduce to scalars as discussed above. Then all possible vector structures can be expressed as
\begin{equation}
\begin{split}
\Gamma^\mu \sim& \{V^\mu_S,V^\mu_T,V^\mu_SS_T\}\\
        =&\left\{\bar P^\mu,\Delta^\mu,n^\mu,\sigma^{\mu\nu} \Delta_{\nu},\sigma^{\mu\nu} n_{\nu},\bar P^\mu\sigma^{\alpha\beta} n_{\alpha}\Delta_{\beta},\Delta^\mu\sigma^{\alpha\beta} n_{\alpha}\Delta_{\beta},n^\mu\sigma^{\alpha\beta} n_{\alpha}\Delta_{\beta}\right\}\ ,
\end{split}
\end{equation}
where each term will be associated with an independent scalar function coefficient when we use this basis to parameterize GPDs.

%
Though more complicated, the $\bar u(P',S')\Gamma^{\mu\nu} u(P,S)$ can be parameterized similarly. For the Dirac scalar contributions, one has 
\begin{equation}
   \Gamma^{\mu\nu}\text{ with Dirac scalar} \sim \{V^\mu_SV^\nu_S,g^{\mu\nu}\}\ ,
\end{equation}
and the Dirac tensor contributions are
\begin{equation}
   \Gamma^{\mu\nu}\text{ with Dirac tensor} \sim \{g^{\mu\nu} S_T,\sigma^{\mu\nu},V^\mu_SV^\nu_S S_T,V^\mu_SV^\nu_T,V^\mu_TV^\nu_S\}\ .
\end{equation}
Consequently, $\Gamma^{\mu\nu}$ can in general be expressed in the basis 
\begin{equation}
\label{Gmunuexp}
   \Gamma^{\mu\nu} \sim \left\{g^{\mu\nu},g^{\mu\nu} S_T,\sigma^{\mu\nu},V^\mu_SV^\nu_S,V^\mu_SV^\nu_S S_T,V^\mu_SV^\nu_T,V^\mu_TV^\nu_S\right\}\ .
\end{equation}
It can be checked that there are 33 independent combinations above and one of them needs to be eliminated in order to be consistent with the form factors counting which will be easier to show later with a look at reparameterization.

\subsection{Reparameterization}

The last thing is to rearrange the parameterization for future convenience. Since one has in general the expansion
\begin{equation}
   V^\mu= p^\mu (V\cdot n)+n^\mu (V\cdot p)+V_\perp^\mu\ ,
\end{equation}
where $V^\mu_\perp$ is non-zero only for $\mu$ in transverse direction, one can always choose a new basis for $V_{S}^\mu$ and $V^{\mu}_T$ as
\begin{align}
    V_S^{\mu}=\{\bar P_+^\mu,\Delta_\perp^\mu,n_-^\mu\} \;\;\;\;\text{and}\;\;\;\;
    V_T^{\mu}=\{\sigma_\perp^{\mu\nu} \Delta_{\nu},\sigma_\perp^{\mu\nu} n_{\nu}\}\ ,
\end{align}
where again $\perp$ means they are non-zero only for $\mu$ in transverse direction. Another useful relation is the Gordon identity eq.~(\ref{gordon1}) projected to $n_\alpha$ which gives
\begin{align}
2 M \bar{u}\left(P',S'\right) \slashed n u(P,S) &=\bar{u}\left(P',S'\right)\left[2+i \sigma^{\alpha \beta}n_\alpha \Delta_{\beta}\right] u(P,S)\ ,
\end{align}
and this indicates that one could use $\{I,\slashed n\}$ instead of $\{I,\sigma^{\alpha \beta}n_\alpha \Delta_{\beta}\}$ as a basis for scalars. Then the following parameterization for $\Gamma^\mu$ as
\begin{equation}
\begin{split}
\Gamma^\mu=&\gamma_{+}^\mu H(x)  + \frac{i\sigma_{+}^{\mu\alpha} \Delta_\alpha}{2M} E(x) \\
&+\frac{i\Delta^\mu_{\perp}}{M}G_{1}(x)+i\Delta^\mu_\perp\slashed n G_{2}(x)+ \frac{i\sigma_{\perp}^{\mu\alpha}\Delta_\alpha}{2M}G_{3}(x)\\
&+  \sigma_\perp^{\mu\nu}n_\nu G_{4}(x)\\
&+  n_-^\mu F_{1}(x) +  n_-^\mu \slashed n F_{2}(x)\ ,
\end{split}
\end{equation}
can be shown to be equivalent, where $x$ here is short for all possible scalars in the system. It should be clear that the first line can be expressed by $P^\mu_+$ and $P^\mu_+ \slashed n$ with the help of the Gordon identities, but it is rewritten to be consistent with the other parameterization. 

For the tensor combination $\bar u(P') \Gamma^{\mu\nu} u(P)$, one has
\begin{equation}
\begin{split}
\Gamma^{\mu\nu}=&A_1(x) g^{\mu\nu}_\perp +A_2(x) g^{\mu\nu}_\perp \slashed n+A_3(x)\sigma^{\mu\nu}_\perp+A_4(x) P^\mu_+ P^\nu_++A_5(x) n^\mu_- n^\nu_-\\
+&A_6(x) \Delta^\mu_\perp \Delta^\nu_\perp+A_7(x) P^\mu_+ n^\nu_-+A_8(x) n^\mu_-P^\nu_+ +A_9(x) P^\mu_+ \Delta^\nu_\perp+A_{10}(x) \Delta^\mu_\perp P^\nu_+ \\
+&A_{11}(x)n^\mu_- \Delta^\nu_\perp+A_{12}(x) \Delta^\mu_\perp n^\nu_-+A_{13}(x) P^\mu_+ P^\nu_+\slashed{n}+\cdots\ .
\end{split}
\end{equation}
By counting the independent combinations in eq.~(\ref{Gmunuexp}), there should be $33=3+3\times3+3\times3+2\times3+3\times2$ independent parameters with 3 elements in $V_{S}$ and $2$ elements in $V_{T}$. However, it can also be shown that 
\begin{equation}
\Delta^{[\mu}_{\perp} \sigma_\perp^{\nu] \rho} \Delta_\rho \in \left\{  \sigma_\perp^{\mu\nu},\Delta_\perp^{[\mu} \sigma_\perp^{\nu] \rho}n_\rho\right\}\ ,
\end{equation}
where $[\mu\nu]$ in the indices means totally anti-symmetrization, so one of the three terms above can be eliminated and only 32 independent parameters left.

\subsection{Another parameterization}
\label{otherparam}

Apparently, the above parameterization is not the only choice. Here another choice of the parameterization will be discussed where the pseudo-vector Dirac bilinear $\bar u \gamma^\mu \gamma_5 u$ is used. More specifically, one has
\begin{equation}
\label{axialvector}
\bar u(P',S') \Gamma^{\mu\nu}u(P,S)\sim \bar u(P',S') \epsilon^{\nu\rho}_{\perp}\Delta_\rho \gamma^{\mu}\gamma_5 u(P,S)+\cdots\ ,
\end{equation}
with
\begin{equation}
\epsilon^{\mu\nu}_\perp\equiv \epsilon^{\alpha\beta\mu\nu}n_\alpha \bar P_\beta\ .
\end{equation}
Given the set of identities at the beginning of this section, the following identity can be derived
\begin{equation}
\bar u(P',S')  \gamma^{\mu}\gamma_5 u(P,S)= \bar u(P',S') \left(\frac{1}{2}\epsilon^{\mu\nu\alpha\beta}\bar P_\nu \sigma_{\alpha\beta}-\frac{1}{8}\Delta^\mu \epsilon^{\alpha\beta\rho\lambda}n_\alpha \Delta_\beta \sigma_{\rho\lambda}\right) u(P,S)\ .
\end{equation}
Following Levi-cevita identity can further simplify it
\begin{equation}
\epsilon^{\mu_1\mu_2\mu_3\mu_4}\epsilon_{\nu_1\nu_2\nu_3\nu_4}=4!\delta^{\mu_1}_{[\nu_1}\delta^{\mu_2}_{\nu_2}\delta^{\mu_3}_{\nu_3}\delta^{\mu_4}_{\nu_4]}\ .
\end{equation}
As a result, it can be shown that the combination in eq.~(\ref{axialvector}) reduces to Dirac bilinears with Dirac scalars and tensors and can be made  coincide with our parameterization. One term that corresponds to the Dirac structure of the coefficient $G_3(x)$ will be of interest
\begin{equation}
\label{axialvectorpara}
\bar u(P',S') i\epsilon^{\nu\rho}_{\perp}\Delta_\rho \gamma^{\mu}\gamma_5 u(P,S)= \bar u(P',S') \left(-i\bar P^\mu \sigma^{\nu\alpha}\Delta_\alpha+\text{other terms}\right) u(P,S)\ .
\end{equation}
It corresponds to the expression in ref. \cite{Ji20132} for instance,  where the coefficient of eq.~(\ref{axialvector}) is used for the AM sum rule which works great since it gives the same structure to the first order of $\Delta$ in the forward limit. However it does not appear to be a good choice for a general parameterization. On the one hand it is made of Dirac pseudo-vectors, and it will be harder to show its completeness when both Dirac pseudo-vectors and Dirac vectors are used in the parameterization. On the other hand, this structure vanishes in the forward limit $\Delta\to 0$, so one should be careful if this can successfully parameterize all of the Dirac structure. Actually it fails to give the PDF for some 3-field correlation when taking the forward limit.

\bibliographystyle{jhep}
\bibliography{refs.bib}
\end{document}